\documentclass[useAMS,usenatbib]{mn2e}

\usepackage{graphicx}
\usepackage{footmisc}
\usepackage{color} 
\usepackage{rotating}
\usepackage{amssymb}
\usepackage{multirow} 
\usepackage{booktabs} 

\newcommand{\lum}{{\rm erg\,s^{-1}}}
\newcommand{\flux}{{\rm erg\,s^{-1}cm^{-2}}}
\newcommand{\kms}{{\rm km\,s^{-1}}}

\newcommand{\red}{$z$}

\newcommand{\aj}{AJ}
\newcommand{\apj}{ApJ}

\newcommand{\mnras}{MNRAS}

\newcommand{\xmm}{\textit{XMM-Newton }}
\newcommand{\nxmm}{\textit{XMM-Newton}}
\newcommand{\chandra}{\textit{Chandra }}

\newcommand{\sgm}{$\sigma$ }                      %
\newcommand{\asec}{$\arcsec$}                      %
\newcommand{\mic}{$\mu$m }                      %
\newcommand{\nmic}{$\mu$m}                      %
\newcommand{\wise}{$\rm WISE$ }                      %
\newcommand{\nwise}{$\rm WISE$}                      %
\newcommand{\sn}{SNR}                      %
\newcommand{\buxs}{$\rm BUXS$ }
\newcommand{\nbuxs}{$\rm BUXS$}
\newcommand{\spitzer}{{\it Spitzer} }
\newcommand{\iras}{{\it IRAS}}
\newcommand{\oiii}{${\rm [O{\,III}]}$}
\newcommand{\loiii}{L$_{\rm [O{\,III}]}$}

\newcommand{\colY}{{\tt log}(${\tt {\it f}_{4.6}/{\it f}_{3.4}}$)}
\newcommand{\colX}{{\tt log}(${\tt {\it f}_{12}/{\it f}_{4.6}}$)}

\title[Obscured AGN selection with \wise]{Uncovering obscured luminous AGN with \wise}
\author[Mateos et al.] {S. Mateos$^{1}$\thanks{E-mail:
    mateos@ifca.unican.es}, A. Alonso-Herrero$^{1}$\thanks{Augusto G. Linares Senior Research Fellow}, F. J. Carrera$^{1}$, A. Blain$^{2}$, P. Severgnini$^{3}$
\newauthor
A. Caccianiga$^{3}$, A. Ruiz$^{4}$
\smallskip \\
\footnotesize
$^{1}$ Instituto de F\'isica de Cantabria (CSIC-Universidad de Cantabria), 39005, Santander, Spain\\
$^{2}$ Physics and Astronomy, University of Leicester, University Road, Leicester LE1 7RH, UK\\
$^{3}$ INAF-Osservatorio Astronomico di Brera, via Brera 28, 20121 Milano, Italy  \\
$^{4}$ Inter University Centre for Astronomy and Astrophysics (IUCAA), Post Bag 4, Ganeshkhind, Pune 411 007, India
}
\begin{document}

\date{Accepted 2013 May 29.  Received 2013 May 20; in original form 2013 March 21}

\pagerange{\pageref{firstpage}--\pageref{lastpage}} \pubyear{2012}

\maketitle

\label{firstpage}

\begin{abstract}
\citet{mateos12} presented a highly reliable and efficient
mid-infrared (MIR) colour-based selection technique for luminous
active galactic nuclei (AGN) using the {\it Wide-field Infrared Survey
  Explorer} (\nwise) survey. Here we evaluate the effectiveness of
this technique to identify obscured AGN missed in X-ray surveys. To do
so we study the \wise properties of AGN independently selected in hard
X-ray and optical surveys. We use the largest catalogue of 887
\oiii\,$\lambda$5007-selected type 2 quasars (QSO2s) at \red$\leq$0.83
in the literature from the Sloan Digital Sky Survey, and the 258 hard
($>$4.5 keV) X-ray-selected AGN from the Bright Ultrahard \xmm Survey
($\rm BUXS$). The fraction of SDSS QSO2s in our infrared AGN selection
region (wedge) increases with the AGN luminosity, reaching ${\rm
  66.1_{-4.7}^{+4.5}}\%$ at the highest \oiii{} luminosities in the
sample. This fraction is substantially lower than for the \buxs type 1
AGN (${\rm 96.1_{-6.3}^{+3.0}}\%$), but consistent, within the
uncertainties, with that for the \buxs type 2 AGN (${\rm
  75.0_{-19.1}^{+14.1}}\%$) with the same luminosity. The SDSS QSO2s
appear to reside in more luminous (massive) hosts than the \buxs AGN,
due to the tight magnitude limits applied in the SDSS spectroscopic
target selection. Since host galaxy dilution can reduce substantially
the effectiveness of MIR-based techniques, this may explain the lower
fraction of SDSS QSO2s in the \wise AGN wedge. The fraction of SDSS
QSO2s identified as Compton-thick candidates that fall in the wedge is
consistent with the fraction of all SDSS QSO2s in that zone. At the
AGN luminosities involved in the comparison, Compton-thick and
Compton-thin SDSS QSO2s have similar WISE colour distributions. We
conclude that at high luminosities and \red$\lesssim$1 our MIR
technique is very effective at identifying both Compton-thin and
Compton-thick AGN.

 \end{abstract}


\section{Introduction}
Providing a complete census of the obscured active galactic nuclei
(AGN) population is crucial to fully understand the cosmological
growth of supermassive black holes (SMBH) and to reveal the physical
nature of the SMBH-galaxy co-evolution (e.g. \citealt{magorrian98};
\citealt{tremaine02}).  A population of heavily obscured AGN at
cosmological distances is required to explain the shape of the Cosmic
X-ray background spectrum as shown by the latest AGN population
synthesis models (\citealt{gilli07}; \citealt{treister09};
\citealt{ballantyne11}; \citealt{shi12}). Deep hard X-ray surveys with
\xmm and \chandra (up to $\sim$10 keV) have provided the most complete
and unbiased samples of AGN spanning a broad range of luminosities and
redshifts (e.g. \citealt{mainieri02}; \citealt{alexander03};
\citealt{brandt05}; \citealt{mateos05}; \citealt{tozzi06};
\citealt{brandt10}; \citealt{xue11}). However, there is strong
observational evidence that even the deepest X-ray surveys conducted
to date are incomplete for AGN with heavy obscuration (line of sight
neutral Hydrogen column densities ${\rm N_H>10^{23}\,cm^{-2}}$) and
they miss many of the objects with Compton-thick (${\rm N_H\gtrsim
  1.5\times10^{24}\,cm^{-2}}$) obscuration (\citealt{ceca08};
\citealt{comastri11}; \citealt{iwasawa12}). Compton-thick AGN could
represent a fraction as large as $\sim$20-30\% of the entire AGN
population (e.g. \citealt{worsley05}; \citealt{burlon11}).

In this context surveys at mid-infrared (MIR) wavelengths
($\gtrsim$5\,\nmic) are much less affected by extinction. Since most
of the absorbed AGN energy is re-emitted in the MIR, surveys at these
wavelengths can potentially recover the elusive obscured accretion
missed by X-ray surveys (e.g. \citealt{lacy04}; \citealt{stern05};
\citealt{alonso06}; \citealt{fiore08}; \citealt{georgantopoulos08};
\citealt{donley12}; \citealt{severgnini12}).

The {\it Wide-field Infrared Survey Explorer} (\nwise;
\citealt{wright10}) has completed the first sensitive survey of the
entire sky at 3.4, 4.6, 12, and 22\,\nmic\, providing an unprecedented
dataset at MIR wavelengths. AGN population studies with \wise are
starting to fill the gap between local/deep MIR surveys with
\iras/\spitzer in regions of the parameter space poorly sampled. MIR
selection techniques rely on identifying objects with red power-law
SEDs, the characteristic signature of hot dust heated by the intense
radiation field of the AGN (\citealt{alonso06}). Several works in the
literature have already demonstrated that using \wise colours alone it
is possible to separate stars and normal/star-forming galaxies from
luminous AGN (\citealt{assef12}; \citealt{mateos12};
\citealt{stern12}; \citealt{yan12}).

In \citet[hereafter M12]{mateos12} we presented an MIR colour-based
selection of luminous AGN candidates using the 3.4, 4.6, and 12\,\mic
bands of \nwise. The reliability and effectiveness of the technique
were evaluated using the 258 hard ($>$4.5 keV) X-ray-selected AGN from
the Bright Ultrahard \xmm Survey ({\rm BUXS}; M12). Surveys such as
\buxs are the most efficient and complete way of tracing the accretion
phenomenon in the Universe at column densities approaching the
Compton-thick limit (e.g. \citealt{caccianiga04};
\citealt{ceca08}). We demonstrated that our technique, aimed to
identify objects with red MIR spectral energy distributions (SEDs), is
highly reliable and very effective for the detection of (X-ray)
luminous AGN. At ${\rm L_{2-10\,keV}}$ in the range
$10^{44}$-$10^{45}\,\lum$, 96.1$_{-6.3}^{+3.0}$\% and
75.0$_{-19.1}^{+14.1}$\% of the \buxs type 1 and type 2 AGN meet the
selection, respectively. Interestingly, we found that hard X-ray
surveys preferentially miss the \wise AGN candidates with the reddest
MIR colours. Objects such as these are the best candidates to account,
at least in part, for the obscured/absorbed luminous AGN largely
missed by X-ray surveys. The hard X-ray signal obtained from stacking
analyses of X-ray non-detected infrared AGN candidates suggests that
in these objects the nuclear emission is obscured by
heavy/Compton-thick material (e.g. \citealt{donley12}).

However, there is still little knowledge on the nature and properties
of the AGN populations identified with \wise and, more importantly,
the impact of heavy nuclear obscuration on the effectiveness of \wise
selection techniques, even at high luminosities. To get insight into
these issues we have evaluated the effectiveness of our MIR selection
technique for the largest catalogue of 887 type 2 quasars (QSO2s) at
\red$\lesssim$0.83 in the literature from the Sloan Digital Sky Survey
(SDSS; \citealt{reyes08}). The objects were selected based on their
high \oiii\,$\lambda$5007 line luminosities, \loiii${\rm \geq10^{8.3}
  L_\odot}$. As \oiii\, fluxes are largely unaffected by AGN nuclear
absorption/obscuration, this sample of luminous QSO2s should not be
biased against heavily obscured objects. Therefore, we can use this
dataset to evaluate whether AGN selection techniques with \wise are
able to identify luminous obscured AGN at \red$\lesssim$1 missed in
X-ray surveys. Furthermore, to get insight into the nature of the
\wise AGN candidates, especially those objects with red infrared SEDs
not detected in X-rays, we studied the properties (infrared colour
distributions and median rest-frame UV-to-MIR SEDs) of the AGN
independently selected in X-ray and optical surveys using the data
from \buxs and the SDSS QSO2 catalogue.

This paper is organized as follows. Section 2 briefly summarizes the
data sets. In Section 3 we determine the effectiveness of our infrared
selection technique for the full sample of SDSS QSO2s and for those
objects identified in the literature as Compton-thick candidates, and
we compare it with the results obtained for the X-ray selected AGN in
the \buxs survey. In Section 4 we compare the distributions of \wise
colours for the different classes of AGN and we compute and present
the properties of their median SEDs. In Section 5 we discuss the main
results of our analysis. The results are summarized in Section
6. Throughout this paper errors are 90\% confidence for a single
parameter ($\Delta\chi^2$=2.71) unless otherwise stated. We estimated
the most probable value for the fractions using a Bayesian approach
and the binomial distribution from \citet{wall08} and the quoted
errors are the narrowest interval that includes the mode and
encompasses 90\% of the probability. The terms ``red" and ``blue"
refer to increasing or decreasing fluxes in $\nu f_\nu$ at longer
wavelengths, respectively. We adopt the concordance cosmology,
$\Omega_M=0.3$, $\Omega_\Lambda=0.7$, and $H_0={\rm
  70\,km\,s^{-1}\,Mpc^{-1}}$.

\section{Sample description}
\label{sample}

\subsection{SDSS catalogue of \oiii\,$\lambda$5007 QSO2s}
\label{qso2_cat}
Here we use the largest catalogue of type 2 quasars in the literature
derived from the SDSS (\citealt{stoughton02}; \citealt{york00})
spectroscopic database (\citealt{reyes08}; hereafter R08). The
catalogue includes objects with UV/optical narrow emission lines only
and with line ratios characteristic of non-stellar ionizing
radiation. As shown by \citet{villar08}, standard AGN photoionization
models can reproduce successfully the main emission-line ratio
diagnostic diagrams at least at high \oiii {} luminosities. The
  R08 catalogue includes only objects with \loiii${\rm \geq10^{8.3}
    L_\odot}$ and \red$<$0.83 to ensure that the \oiii\,$\lambda$5007
  line is present in the SDSS spectra. The R08 catalogue includes
both Gaussian and non-parametric \oiii\, line measurements, the latter
(used to define the luminosity cut) being systematically larger by 5\%
on average. In this paper \loiii{} refers to luminosities estimated
from Gaussian fits. The catalogue contains 887 type 2 quasars detected
over a survey area of $\approx$6293 deg$^2$. From now on, we will
refer to this source sample as the SDSS QSO2s.

The objects in R08 were selected independently of their X-ray
properties with a selection that should be unaffected by nuclear
obscuration. Only a small subset of the SDSS QSO2s have X-ray
information, which is necessary to directly measure the absorption
along the line of sight. Detailed analyses of their X-ray properties
suggest that, at least at the highest \oiii\, luminosities, most SDSS
QSO2s are heavily absorbed and an important fraction might be
Compton-thick (\citealt{ptak06}; \citealt{lamassa09};
\citealt{vignali10}; \citealt{jia12}). Similar results are found for
\oiii\,$\lambda$5007-selected AGN at lower luminosities
\citep{goulding11}. Thus, we can confidently use the SDSS QSO2
catalogue to evaluate the effectiveness of the infrared AGN wedge to
select luminous absorbed AGN at \red$\lesssim$1 in a way that is
relatively free from the potential selection effects of X-ray
absorption.

We caution however, that the objects in the R08 catalogue were
selected from different SDSS spectroscopic target selection algorithms
which implies that the selection function is different across the
luminosity-\red{} range of the sample. At \red$\leq$0.3 (and lower
\oiii{} luminosities) most SDSS QSO2s were targeted by the Main Galaxy
algorithm (based on optical morphology i.e. resolved objects) while at
\red$>$0.3 (and higher \oiii{} luminosities) the SDSS QSO2s were
mainly targeted by the Low-\red{} and High-\red{} QSO (non-stellar
colour selection) and Serendipity FIRST (radio properties) algorithms
  (see Fig. 3 in R08). We will discuss how this property of the SDSS
  QSO2 catalogue might affect our results. A summary of the SDSS
  spectroscopic target selection algorithms is given in R08 while for
  more detailed information see \citet{richards02} and
  \citet{strauss02}.

\subsection{The Bright Ultra-hard \xmm Survey} \label{buxs}
\buxs is a complete flux-limited sample of bright (${\rm {\it
    f}_{4.5-10\,keV} > 6\,x\,10^{-14}\,\flux}$) ``ultra-hard'' (4.5-10
keV) X-ray selected sources detected over a total sky area of 44.43
deg$^2$.  \buxs is based on a subset of 381 high galactic latitude
($|b|>20\deg$) observations from the second \xmm serendipitous source
catalogue (2XMM; \citealt{watson09}). These observations, with
effective exposure times in the range 10-100 ks, were used to derive
high precision X-ray extragalactic source count distributions at
intermediate fluxes (\citealt{mateos08}). At the time of writing the
spectroscopic identification completeness is 97.3\%. Of the 258 \buxs
sources, 145 objects (56.2\%) are identified as type 1 AGN (showing
UV/optical emission line velocity widths $\geq$1500 $\kms$) and 106
(41.1\%) as type 2 AGN (showing UV/optical emission line velocity
widths $<$1500 $\kms$ or a galaxy spectrum with no emission lines)
while 7 sources (2.7\%) remain unidentified. \buxs covers 4 decades in
X-ray luminosity (${\rm \sim10^{42}-10^{46}\,\lum}$) and extends up to
\red$\sim$2. X-ray luminosities are computed in the ``standard'' 2-10
keV rest-frame energy band. They were derived from a detailed analysis
of the X-ray spectra of the objects and they are corrected for
intrinsic absorption.

\begin{table*}
  \begin{center}
    \caption{Summary of the properties of the different source samples used in this study.}
    \label{tab0}
    \begin{tabular}{cccccccc}
      \hline
      \hline
      Sample & ${\rm N_{tot}}$ & ${\rm N_{SNR\ge5}}$ &  ${\rm N_{wedge}}$ & $\Omega$ & $f_{12\mu {\rm m}}$ & $z$ & log\,(L$_{2-10\,{\rm keV}}$)  \\
       & &  & & (deg$^2$) &  (mJy) & & (erg/s)\\
      (1) & (2) & (3) & (4) & (5) & (6) & (7) & (8)\\
      \hline
      SDSS QSO2s    &  887   &   766      &   408     &   6293   &  3.21   & 0.26 & 43.88  \\  
      BUXS\,type\,1 &  145   &   114      &   105     &   44.43  &  1.32   & 0.64 & 44.34  \\  
      BUXS\,type\,2 &  106   &    81      &    38     &   44.43  &  1.71   & 0.23 & 43.38  \\  
      BUXS\,WISE    & 448909 &   25206    &   2755    &   44.43  &  0.82   &  -    &  - \\  
      BUXS\,WISE+X  & 6377   &   1659     &   1062    &   44.43  &  0.96   &  -    &  - \\  
      \hline
    \end{tabular} \\
Column 1: Source sample; column 2: total number of objects in sample;
column 3: number of objects detected with \sn$\ge$5 in all the first
three bands of \nwise; column 4: number of objects in the infrared AGN
wedge; column 5: solid angle of the survey; columns 6, 7 and 8: median
12\,\mic flux, \red{} and 2-10 keV luminosity for those objects with
detection \sn$\ge$5 in all the first three bands of \nwise. For
comparison, the median 12\,\mic flux of the \wise AGN candidates in
the infrared wedge not detected in X-rays is 0.75 mJy. For the SDSS
QSO2s the luminosities were derived using the empirical relation
between hard X-ray emission and \oiii\,$\lambda$5007 luminosity from
\citet{jin12} (see Sec.~\ref{comparison}). For the \buxs AGN, the
luminosities were computed from the analysis of their X-ray spectra
and they are corrected for Galactic and intrinsic absorption.
  \end{center}
\end{table*}

\subsection{\wise infrared data}
\label{wise_data}
We use here the most recent publicly available All-Sky Data Release
that covers $>$99\% of the sky (March 2012;
\citealt{cutri12}\footnote{http://wise2.ipac.caltech.edu/docs/release/allsky/}).
\wise estimated 5\sgm point source sensitivities are better than 0.07,
0.1, 0.9, and 5.4 mJy at 3.4, 4.6, 12, and 22\,\nmic, respectively,
although the depth increases with the ecliptic latitude
\citep{jarrett11}.  The angular resolution is (FWHM) 6.1\asec,
6.4\asec, 6.5\asec, and 12\asec\, in the four \wise bands,
respectively. The astrometric precision for sources with
signal-to-noise ratio (hereafter \sn) \sn$>$20 in at least one \wise
band is better than 0.15\asec\, \citep{wright10}.

We computed flux densities using profile fitting photometry and the
magnitude zero points of the Vega system that correspond to a
power-law spectrum ($f_\nu\propto\nu^{\alpha}$) with spectral index
$\alpha$=-1 \citep{wright10}: $F_\nu$ (iso)=309.124 Jy, 171.641 Jy,
30.988 Jy, and 8.346 Jy for 3.4, 4.6, 12, and 22\,\nmic,
respectively. Using the flux correction factors that correspond to
constant power-law spectra the difference in the computed flux
densities would be less than 0.2\% at 3.4, 4.6, and 22\,\mic and
$\sim$2\% at 12\,\nmic. We added a 1.5\% uncertainty to the catalogued
flux errors in all bands to account for the overall systematic
uncertainty from the Vega spectrum in the flux zero-points. To account
for the existing discrepancy between the red and blue calibrators used
for the conversion from magnitudes to Janskys an additional 10\%
uncertainty was added to the 12\,\mic and 22\,\mic fluxes
\citep{wright10}. Throughout this paper we use monochromatic MIR flux
densities ($f_\nu$) in Janskys, unless otherwise specified.

\begin{figure}
  \centering
  \begin{tabular}{cc}
    \hspace*{-0.7cm}\includegraphics[angle=90,width=0.49\textwidth]{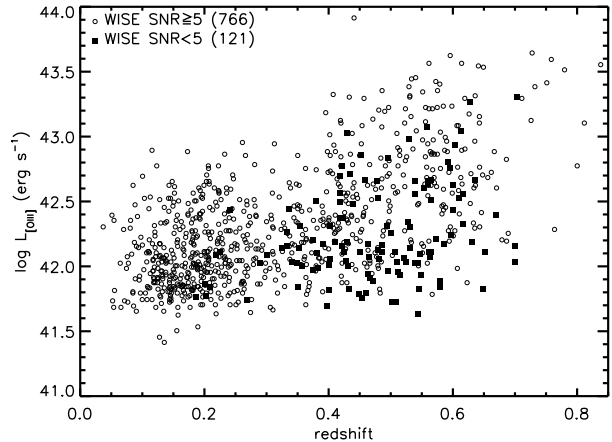}\\
  \end{tabular}
\vspace{+0.05in}
  \caption{Distribution of the \oiii\,$\lambda$5007 line luminosity
    (in logarithmic units) versus \red{} for the SDSS QSO2s in the R08
    catalogue. Objects detected with \sn$\geq$5 in the three shorter
    wavelength bands of \wise are indicated with open circles. Filled
    squares indicate the objects without \wise counterparts and those
    with \sn$<$5 in any of the three shorter wavelength bands of
    \nwise\,(17 and 104, respectively).}
  \label{fig1}
\end{figure}

\begin{figure*}
  \centering
  \begin{tabular}{cc}
    \includegraphics[angle=90,width=0.46\textwidth]{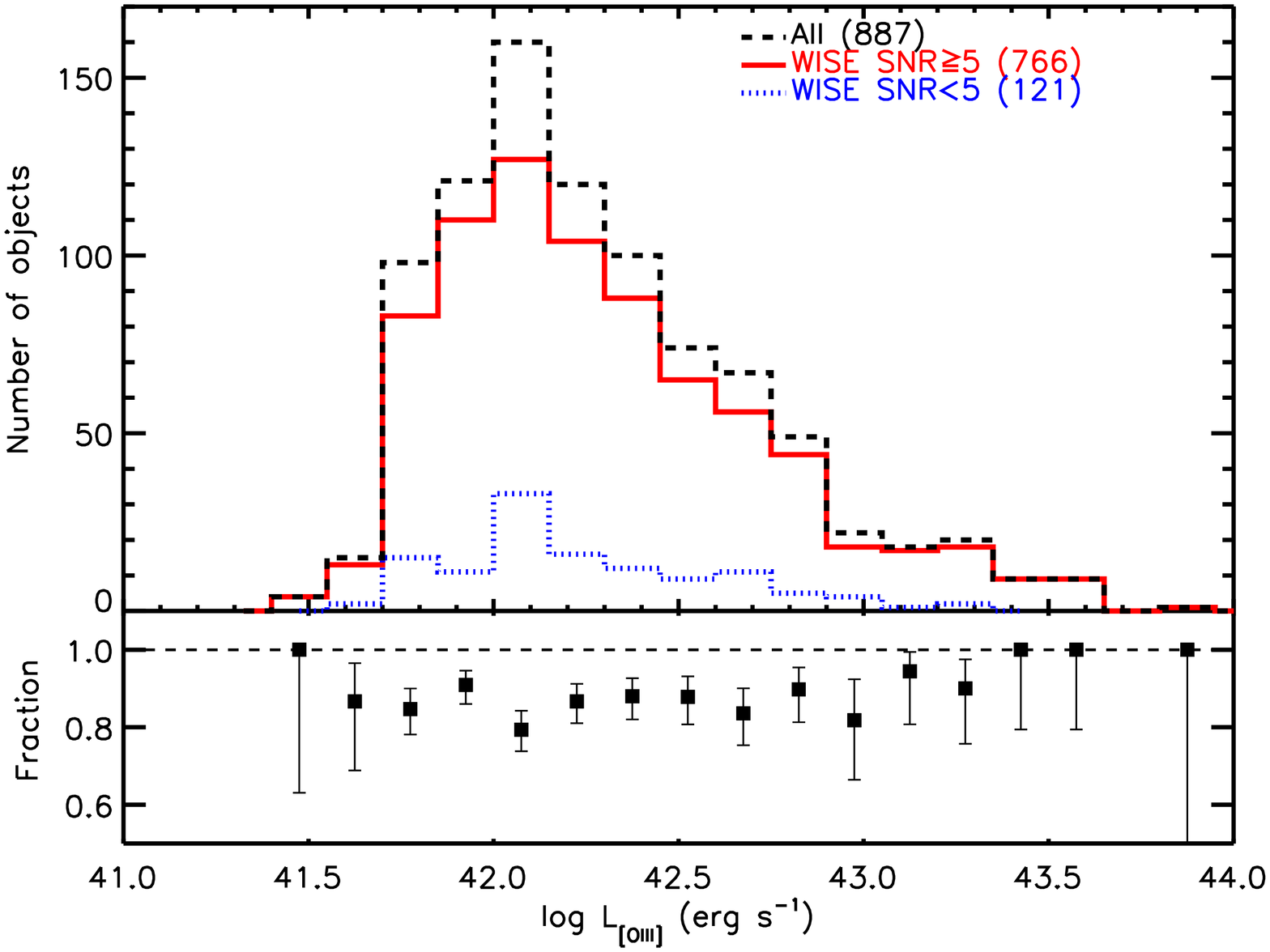}
    \includegraphics[angle=90,width=0.46\textwidth]{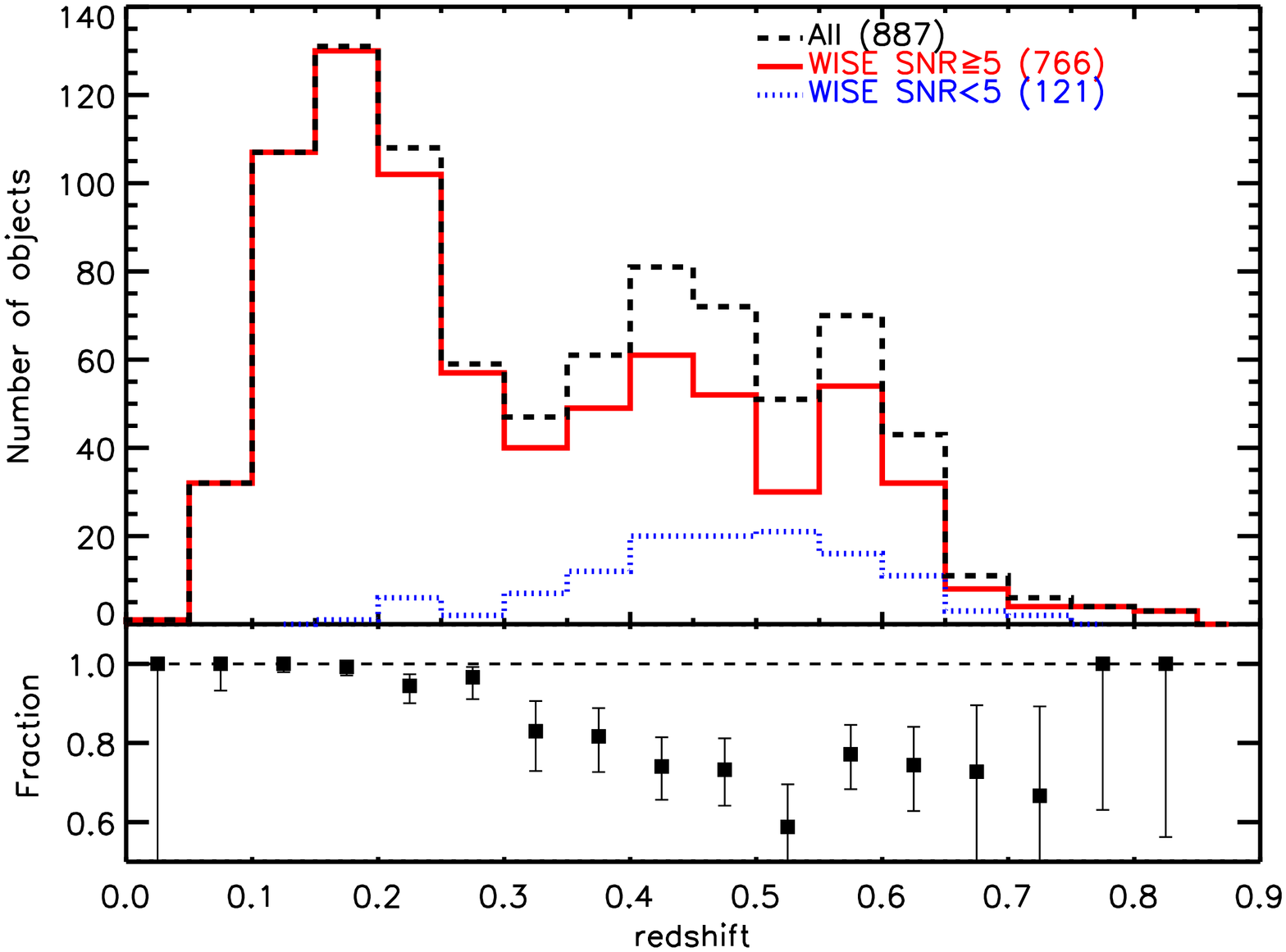}\\
  \end{tabular}
\vspace{+0.05in}
  \caption{\oiii\,$\lambda$5007 luminosity (in logarithmic units,
    left) and \red{} (right) distributions of the SDSS QSO2s in
    R08. We show the distributions for the full sample (dashed), those
    objects detected with \sn$\geq$5 at all 3.4, 4.6, and 12\,\mic
    (solid), and those without \wise counterparts or with \sn$<$5 in
    any of the three shorter wavelength bands of \nwise\,(17 and 104,
    respectively). The plots at the bottom indicate the fraction of
    SDSS QSO2s detected with \sn$\geq$5 at all 3.4, 4.6, and 12\,\mic
    as a function of \oiii\,$\lambda$5007 luminosity (left) and \red{}
    (right), respectively.}
  \label{fig2}
\end{figure*}

\subsubsection{Detection of SDSS QSO2s with \wise}
\label{det_completeness}
We cross-matched the SDSS QSO2 and the all-sky WISE catalogues using a
search radius of 2\asec. All but 17 objects have MIR counterparts. We
checked that for two of these sources a blend of two WISE objects
prevents us from identifying a unique MIR counterpart. The remaining
15 sources are not detected with \wise even if we increase the search
radius.

Since in this paper we use the \wise AGN selection technique from M12
(see Sec.~\ref{wedge-def}), which identifies AGN candidates among
\wise objects detected with \sn$\geq$5 at all 3.4, 4.6, and 12\,\nmic,
in the following we use the same infrared detection
criteria. Table~\ref{tab0} summarizes the main properties of the SDSS
QSO2s selected for this study as well as those of the \buxs AGN (see
Sec.~\ref{wise_buxs}). Fig.~\ref{fig1} shows the \oiii\,$\lambda$5007
line luminosity versus \red{} distribution for the SDSS QSO2s with
\sn$\geq$5 in all three (3.4, 4.6, and 12\,\nmic) bands (766) and
without detection or \sn$<$5 in any of those bands (121, of which 17
do not have a \wise counterpart). Fig.~\ref{fig2} compares their
\oiii\,$\lambda$5007 luminosity (left) and \red{} (right)
distributions and indicates the fraction of SDSS QSO2s detected in all
the first three bands of WISE as a function of luminosity and \red{}
(bottom panels). As Fig.~\ref{fig1} shows, we lose preferentially the
least luminous (on the basis of their \oiii\,$\lambda$5007 line
luminosity) SDSS QSO2s at \red$>$0.3. Still, the detection fraction is
high and remains above $\sim$70\% over most of the luminosity-\red{}
range. This ensures that our \wise three-band detection and \sn{}
requirements will not bias any of the results of the paper.

\begin{figure*}
  \centering
  \begin{tabular}{cc}
    \hspace*{-0.7cm}\includegraphics[angle=90,width=0.8\textwidth]{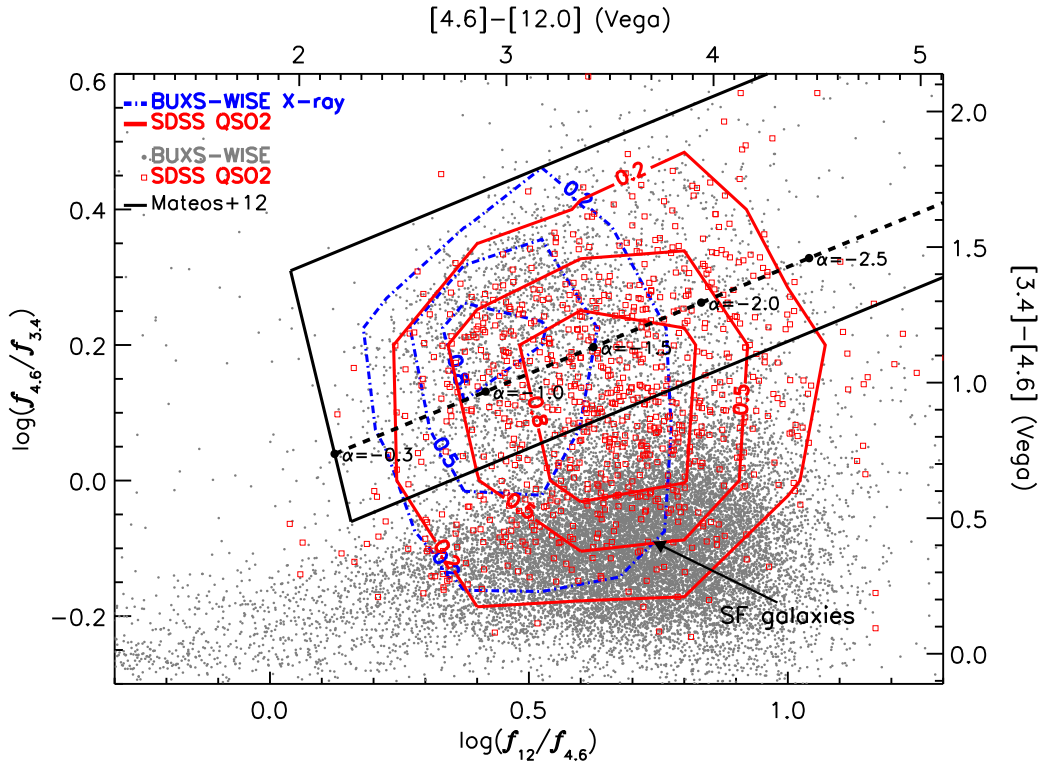}\\
  \end{tabular}
\vspace{+0.15in}
  \caption{\wise MIR colours for sources detected with \sn$\geq$5 at
    all 3.4, 4.6, and 12\,\nmic. Filled circles are all the \wise
    sources in the \buxs survey area, i.e., not applying any of the
    X-ray selection criteria that define the \buxs AGN sample (see
    Sec.~\ref{wise_buxs} and Table~\ref{tab0}). The SDSS QSO2s in R08
    are shown as open squares. The M12 infrared AGN wedge and
    power-law locus (and the values for different spectral index) are
    the thick solid and dashed lines, respectively. The solid and
    dotted-dashed contours indicate the densities (normalized to the
    peak value) of the SDSS QSO2s and all the \wise sources in
      the \buxs area detected at 2-10 keV energies down to the full
      depth of the X-ray data, respectively.}
  \label{fig3}
\end{figure*}

\subsubsection{Matching \wise with \buxs}
\label{wise_buxs}
To reliable identify the \wise counterparts of the \buxs AGN we used
the cross-matching algorithm of \citet{pineau11}, which is based on
the classical likelihood ratio. Full details on the cross-correlation
are given in M12. In total 255 out of 258 objects (98.8\%) have \wise
counterparts (i.e. detection with \sn$\ge$5 in at least one band). The
number of such objects detected with \sn$\geq$5 at all 3.4, 4.6, and
12\,\nmic, is 114 out of 145 (78.6\%) type 1 AGN and 81 out of 106
(76.4\%) type 2 AGN (see Table~\ref{tab0}).

To get insight into the nature of the \wise objects identified as AGN,
in M12 we studied the properties (infrared colours) of all catalogued
\wise sources in the \buxs survey area detected in all the first three
bands of \wise (25206 objects), and of those sources with X-ray
detection at 2-10 keV energies down to the full depth of the X-ray
data (1659 objects). One of the most interesting results from the
  M12 study is that the fraction of \wise objects identified as AGN
  from X-rays decreases substantially at the reddest MIR colours (see
  Fig.~2 in M12). Objects of this sort could account for part of the
  population of absorbed AGN largely missed by hard X-ray
  surveys. Currently only a small fraction of the \wise AGN candidates
  not detected in X-rays in the \buxs area have optical spectroscopic
  identifications ($<$9\%), almost all from SDSS. Clearly these
  identifications are biased towards blue type 1 AGN and therefore
  they are not representative of the underlying AGN population. Thus,
  to get more insight into the nature of the \wise AGN, in this work
  we compare their infrared colours with those of the \buxs type 1 and
  type 2 AGN and the SDSS QSO2s.

Table~\ref{tab0} gives a summary of the properties of the different
source samples derived from the \buxs survey that we use in this
study.

\section{\wise AGN selection for SDSS QSO2} \label{selection}
\subsection{Our infrared AGN selection technique} \label{wedge-def}
In M12 we used the data from the \buxs survey to define an MIR colour
selection of luminous AGN using the three shorter wavelength bands of
\wise (hereafter infrared AGN wedge). The technique, that properly
takes into account the errors in the photometry, can robustly identify
with high completeness X-ray luminous AGN down to a \sn=5 significance
of detection at all 3.4, 4.6, and 12\,\nmic. At ${\rm L_{2-10\,keV}}$
from $10^{44}$ to $10^{45}\,\lum$, 96.1$_{-6.3}^{+3.0}$\% and
75.0$_{-19.1}^{+14.1}$\% of the \buxs type 1 and type 2 AGN meet the
selection, respectively. The number of \wise objects in the \buxs area
in the infrared AGN wedge (both with and without X-ray detection) are
reported in Table~\ref{tab0}.  We note than the overlap of the \buxs
and SDSS QSO2 samples is very small as there are only three sources in
common between the SDSS QSO2s and the \wise sources in \buxs in the
infrared AGN wedge.

\subsection{Effectiveness of the \wise selection as a function of \oiii\, luminosity}
\label{sel_completeness}
The distribution of \wise infrared colours of the SDSS QSO2s is shown
in Fig.~\ref{fig3} (open squares and thick solid line
contours). Fig.~\ref{fig3} also shows the infrared AGN wedge defined
in M12 and the power-law locus for different values of the spectral
index\footnote{Power-law spectrum $f_\nu \propto\nu^\alpha$ with
  spectral index $\alpha$} with solid and dashed lines,
respectively. For comparison we show the colours of the WISE sources
detected in the \buxs fields (small filled circles) and of those
  objects with detection at 2-10 keV X-ray energies (dotted-dashed
  contours). As in M12 we are assuming that an X-ray detection $>$2
  keV is a good tracer of unabsorbed and mildly absorbed AGN activity.

\begin{figure}
  \centering
  \begin{tabular}{cc}
    \hspace{-0.0cm}\includegraphics[angle=90,width=0.47\textwidth]{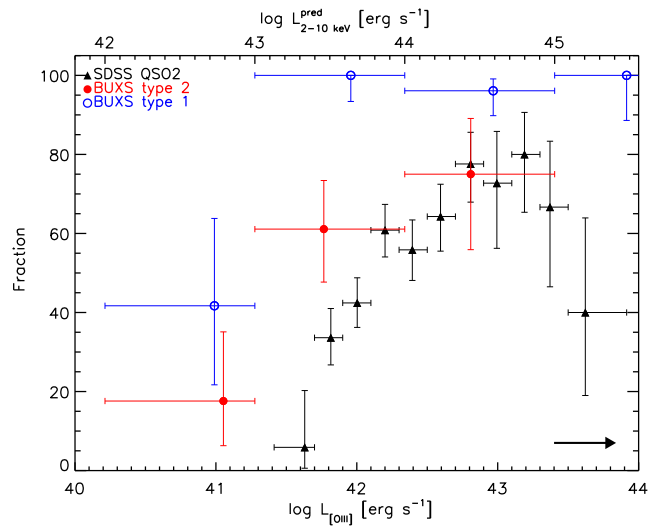}
  \end{tabular}
\vspace{+0.05in}
  \caption{Fraction of SDSS QSO2s in the infrared AGN wedge as a
    function of the \oiii\,$\lambda$5007 line luminosity (in
    logarithmic units; triangles). The symbols indicate the mean
    luminosity of the sources in the bin. The top axis indicates the
    predicted rest-frame 2-10 keV luminosities derived from the
    empirical relation between the X-ray and \oiii\,$\lambda$5007
    luminosities of J12 (see Sec.~\ref{comparison}). The results
    obtained for the \buxs type 1 and type 2 AGN are indicated with
    open and filled circles, respectively. For the \buxs AGN the X-ray
    luminosities were derived from the modeling of their X-ray spectra
    and are corrected for intrinsic absorption. The horizontal arrow
    at the bottom right shows the amplitude of the median extinction
    correction to the \oiii{} line luminosities (see
    Sec.~\ref{comparison}).}
  \label{fig4}
\end{figure}

From Fig.~\ref{fig3} it is clear that a substantial fraction of the
SDSS QSO2s have infrared colours typical of low \red{} star-forming
galaxies (horizontal sequence in the lower-right part of the
diagram). Indeed, out of the 766 SDSS QSO2s with detection in the
three shortest wavelength bands of \wise only 408 (53.3$\pm$3\%) fall
in the infrared AGN wedge. The remainder are in a continuous extension
to bluer \colY{} colours, presumably due to increasing amounts of
contributed flux from starlight in the AGN host galaxy. At low
luminosities and for objects with large extinction at rest-frame
infrared wavelengths, the emission from the host galaxy contaminates
the observed flux preventing the detection of the AGN signatures with
infrared colour and power-law based selection techniques.  In
particular, the shortest wavelengths will be the most contaminated
ones producing a shift towards the lowest-right part of the
diagram. Thus, the effectiveness of any MIR selection strongly depends
on the AGN luminosity (e.g. \citealt{alonso06}; \citealt{eckart10};
\citealt{donley12}; \citealt{mateos12}). Fig.~\ref{fig4} shows the
dependence on luminosity of the fraction of SDSS QSO2s in the infrared
AGN wedge. The fractions were computed in bins of 0.2 in luminosity
and are reported in Table~\ref{tab1}. The infrared selection is, as
expected, a strong function of the AGN luminosity.

There is a decrease in the fraction of SDSS QSO2s in the infrared AGN
wedge at the highest \oiii{} luminosities. We caution however, that at
these luminosities and \red{} the SDSS QSO2 catalogue is far from
being complete or representative of the QSO2 population (see
Fig.~\ref{fig1}). Furthermore, as noted in Sec.~\ref{qso2_cat} careful
attention must be paid when comparing the results for objects from
different SDSS spectroscopic target selections. In Table~\ref{tab1} we
compare the fractions of SDSS QSO2s in the infrared AGN wedge for
objects at \red$\leq$0.3 and \red$>$0.3, respectively. Although the
values are consistent within the uncertainties, at the highest
luminosities they are systematically lower for the SDSS QSO2s at
\red$>$0.3. This effect is most likely associated with the different
selection functions at \red$\leq$0.3 and \red$>$0.3. For example, at
\red$>$0.3 most sources were targeted by the QSO algorithms (see
Sec.~\ref{qso2_cat}) that pre-select objects with non-stellar colours
(blue) in the $ugriz$ broad-band photometry. It is not unexpected that
their blue optical SEDs will extend to the near-infrared, and hence
these objects would easily be missed by infrared selection techniques
targeting AGN with red infrared SEDs. However, we cannot rule out that
the apparent decrease in effectiveness is due in part to the
k-correction (i.e. the host galaxy emission contaminates the \wise
fluxes at the shortest wavelengths).

\begin{table*}
  \begin{center}
    \caption{Luminosity and \red{} dependence of the fraction of SDSS QSO2s in the infrared AGN wedge.}
    \label{tab1}
    \begin{tabular}{ccccccc}
      \hline
      \hline
      ${\rm log(L_{[OIII]})}$ & ${\rm N_{tot}}$ & ${\rm {\it f}_{tot}}$ & ${\rm N_{z\leq 0.3}}$ & ${\rm {\it f}_{z\leq 0.3}}$ & ${\rm N_{z> 0.3}}$ & ${\rm {\it f}_{z> 0.3}}$  \\
      (1) & (2) & (3) & (4) & (5) & (6) & (7)\\
      \hline
      {[}41.4$-$41.7{]} &  17  &  ${\rm 5.9_{-5.3}^{+14.4}}$    &   15  &  ${\rm 6.7_{-6.0}^{+15.9}}$  &    2  & ${\rm 0.0_{}^{+53.6}}$\\  
      {[}41.7$-$41.9{]} & 116  &  ${\rm 33.6_{-6.9}^{+7.4}}$    &  101  &  ${\rm 32.7_{-7.3}^{+7.9}}$  &   15  & ${\rm 40.0_{-18.0}^{+20.1}}$\\   
      {[}41.9$-$42.1{]} & 165  &  ${\rm 42.4_{-6.2}^{+6.3}}$    &  129  &  ${\rm 40.3_{-6.9}^{+7.2}}$  &   36  & ${\rm 50.0_{-13.2}^{+13.2}}$\\   
      {[}42.1$-$42.3{]} & 143  &  ${\rm 60.8_{-6.8}^{+6.5}}$    &   83  &  ${\rm 66.3_{-8.7}^{+8.0}}$  &   60  & ${\rm 53.3_{-10.4}^{+10.2}}$\\   
      {[}42.3$-$42.5{]} & 111  &  ${\rm 55.9_{-7.7}^{+7.6}}$    &   54  &  ${\rm 63.0_{-11.0}^{+10.1}}$ &   57  & ${\rm 49.1_{-10.6}^{+10.7}}$\\   
      {[}42.5$-$42.7{]} &  84  &  ${\rm 64.3_{-8.8}^{+8.2}}$    &   36  &  ${\rm 77.8_{-12.4}^{+9.7}}$  &   48  & ${\rm 54.2_{-11.6}^{+11.3}}$\\    
      {[}42.7$-$42.9{]} &  58  &  ${\rm 77.6_{-9.7}^{+8.0}}$    &   12 &  ${\rm 91.7_{-18.9}^{+7.4}}$   &   47  & ${\rm 74.5_{-11.1}^{+9.3}}$\\   
      {[}42.9$-$43.1{]} &  22  &  ${\rm 72.7_{-16.5}^{+13.1}}$  &    -  &      -                      &   21  & ${\rm 71.4_{-17.0}^{+13.6}}$          \\ 
      {[}43.1$-$43.3{]} &  25  &  ${\rm 80.0_{-14.6}^{+10.6}}$  &    -  &      -                      &   25  & ${\rm 80.0_{-14.6}^{+10.6}}$             \\ 
      {[}43.3$-$43.5{]} &  15  &  ${\rm 66.7_{-20.2}^{+16.7}}$  &    -  &      -                      &   15  & ${\rm 66.7_{-20.2}^{+16.7}}$      \\ 
      {[}43.5$-$44.0{]} &  10  &  ${\rm 40.0_{-21.0}^{+23.9}}$  &    -  &      -                      &   10  & ${\rm 40.0_{-21.0}^{+23.9}}$      \\ 
      \hline
      Total             & 766  &   ${\rm 53.3_{-3.0}^{+3.0}}$    &    430 &  ${\rm 49.8_{-4.0}^{+4.0}}$ &   336  &  ${\rm 57.7_{-4.4}^{+4.4}}$  \\ 
      \hline
    \end{tabular} \\
Column 1: X-ray luminosity range of the bin in units of $\lum$ (logarithmic
units); column 2: number of SDSS QSO2s; column 3: fraction of SDSS
QSO2s in the infrared AGN wedge; column 4: number of SDSS QSO2s at
\red$\leq$0.3; column 5: fraction of SDSS QSO2s at \red$\leq$0.3 in
the infrared AGN wedge; column 6: number of SDSS QSO2s at \red$>$0.3;
column 7: fraction of SDSS QSO2s at \red$>$0.3 in the infrared AGN wedge.

  \end{center}
\end{table*}

\subsection{Comparison with an X-ray selected sample}
\label{comparison}
Most SDSS QSO2s in R08 have not been observed in X-rays. Hence, to
  compare with the results for the AGN in \buxs (overplotted in
  Fig.~\ref{fig4}) we derived intrinsic 2-10 keV rest-frame
  luminosities using the empirical relation between hard X-ray
  emission and \oiii\,$\lambda$5007 luminosity for type 1 AGN from
  \citet{jin12} (hereafter J12),

\begin{eqnarray}
{\rm log(L_{2-10\,keV})=(0.94\pm0.05)\times log(L_{[OIII]})+(4.20\pm2.26)}
\end{eqnarray}

Here we are assuming that the above relation is the same for both type
1 and type 2 AGN, and that observed differences are mainly due to
obscuration/absorption (e.g. \citealt{heckman05}). J12 used a large
sample of SDSS type 1 AGN with high quality \xmm spectroscopy, low
reddening and unobscured X-ray emission. This allowed to constrain
better the empirical relation compared to previous works
(e.g. \citealt{heckman05}; \citealt{lamastra09}; \citealt{trouille10};
\citealt{jin12}). Using the J12 relation we are also accounting for
systematic differences in the \loiii{}/${\rm L_{2-10\,keV}}$
distributions measured for X-ray and optically selected AGN
(e.g. \citealt{netzer06}).

We have not corrected for extinction the \oiii\,$\lambda$5007
luminosities, hence predicted X-ray luminosities will be somewhat
higher for some SDSS QSO2s. For many objects, the SDSS spectral
coverage does not reach the rest-frame wavelengths of the H$\alpha$
line and the \sn{} of higher order Balmer lines is not sufficient to
reliable measure line luminosities. Furthermore, extinctions based on
the H$\alpha$/H$\beta$ and H$\beta$/H$\gamma$ ratios are inconsistent
with standard reddening laws (see R08). R08 derived a median Balmer
line ratio H$\alpha$/H$\beta$=4.25 for 44 SDSS QSO2s with
\loiii$>3.8\times10^{42}\lum$ at \red$<$0.4. Using the
\citet{bassani99} relation to derive extinction-corrected
luminosities, ${\rm log(L_{[OIII]}^c)}$, we obtain ${\rm
  log(L_{[OIII]}^c)}$=${\rm log(L_{[OIII]})}$+0.44, for an intrinsic
Balmer decrement of 3.0. Assuming that all SDSS QSO2s suffer the same
level of extinction, such uncertainty level (indicated by the
horizontal arrow at the bottom right of Fig.~\ref{fig4}) does not
affect our main conclusions.\\

Fig.~\ref{fig4} (see also Table~\ref{tab2}) shows that the fraction of
SDSS QSO2s in the infrared AGN wedge is substantially lower than for
the \buxs type 1 AGN but it is consistent, within the uncertainties,
with that for the \buxs type 2 AGN, especially at the highest
luminosities. However, if we correct for extinction all \oiii{}
luminosities using the median Balmer line ratio derived in R08, the
fraction of SDSS QSO2s turns out to be significantly lower than that
of the \buxs type 2 AGN at ${\rm L_{2-10\,keV}\lesssim10^{44}\,\lum}$
(last column in Table~\ref{tab2}). Even if this result must be
considered with caution, given the large uncertainties related to the
reddening correction, we can conclude that the completeness function
of SDSS QSO2s is lower than that of the \buxs type 2 AGN in the low
luminosity regime.

\begin{table}
  \begin{center}
    \caption{Comparison of the effectiveness of the infrared AGN wedge for the \buxs AGN and SDSS QSO2s.}
    \label{tab2}
    \begin{tabular}{ccccccc}
      \hline
      \hline 
       ${\rm log(L_{2-10\,keV})}$ & ${\rm {\it f}_{type\,1}}$ & ${\rm {\it f}_{type\,2}}$ & ${\rm {\it f}_{QSO2s}}$ & ${\rm {\it f}^{corr}_{QSO2s}}$ \\
      (1) & (2) & (3) & (4) & (5)\\
      \hline
      {[}43$-$44{]} &  ${\rm 100_{-6.6}^{}}$   &  ${\rm 61.1_{-13.4}^{+12.3}}$  & ${\rm 45.5_{-3.8}^{+3.8}}$ & ${\rm 30.1_{-6.2}^{+6.7}}$\\  
      {[}44$-$45{]} &  ${\rm 96.1_{-6.3}^{+3.0}}$ &  ${\rm 75.0_{-19.1}^{+14.1}}$ & ${\rm 66.1_{-4.7}^{+4.5}}$ & ${\rm 56.9_{-3.4}^{+3.4}}$\\  
      \hline
    \end{tabular} \\
Column 1: X-ray luminosity range of the bin in units of $\lum$ (logarithmic
units); columns 2, 3 and 4: fraction of \buxs type 1 and type 2 AGN
and SDSS QSO2s in the infrared AGN wedge, respectively; column 5: fraction of SDSS
QSO2s in the infrared AGN wedge after applying a median extinction
correction to the \oiii{} line luminosities (see Sec.~\ref{comparison}
for details).
  \end{center}
\end{table}

The apparent different fractions of \buxs type 1 and type 2 AGN in the
infrared AGN wedge, especially at low luminosities, could be explained
if all type 2 objects suffer larger extinction at rest-frame
optical/near-infrared wavelengths so that, for a given luminosity,
their observed \wise fluxes are more contaminated by their host
galaxies than those of type 1 AGN. We note however, that the optical
classification of the AGN in \buxs relies on the detection of
UV/optical broad emission lines. Therefore, we cannot rule out the
possibility that some \buxs type 2 AGN might be indeed low luminosity
type 1 AGN where the broad emission lines are diluted by the host
galaxy starlight. This may have some effect on the derived fractions
of type 1 AGN in the infrared AGN wedge at ${\rm
  L_{2-10\,keV}<10^{44}\lum}$. However, such effect should be
negligible at luminosities ${\rm L_{2-10\,keV}>10^{44}\lum}$
(e.g. \citealt{caccianiga07}).

\begin{figure}
  \centering
  \begin{tabular}{cc}
    \includegraphics[angle=90,width=0.45\textwidth]{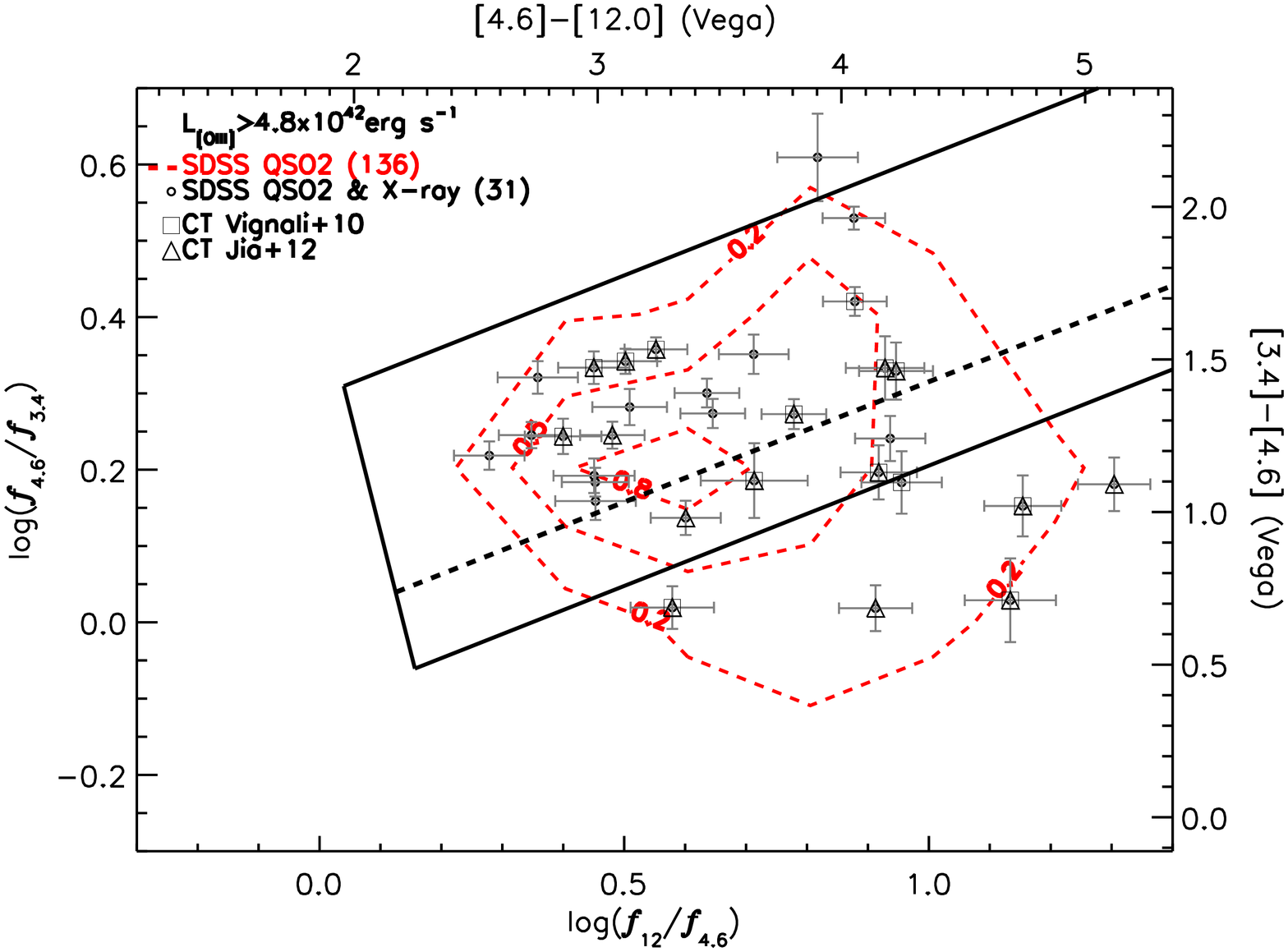}\\
  \end{tabular}
\vspace{+0.15in}
  \caption{MIR colours for the SDSS QSO2s with \wise detections
      with \sn$\geq$5 at all 3.4, 4.6, and 12\,\nmic{},
      \loiii$\geq$4.8$\times$$10^{42}\,\lum$ and X-ray follow-up with
      either \chandra or \nxmm (open circles). The error bars include
      both the \wise catalogued flux errors and systematic
      uncertainties as described in
      Sec.~\ref{wise_data}. Compton-thick candidates identified in the
      literature from the studies of \citet{vignali10} and
      \citet{jia12} are indicated with open squares and triangles,
      respectively. Our AGN selection wedge and power-law locus are
      the thick solid and dashed black lines, respectively. The dashed
      contours indicate the density (normalized to the peak value) of
      all SDSS QSO2s with \loiii$\geq$4.8$\times$$10^{42}\,\lum$.}
  \label{fig4b}
\end{figure}

We would expect different fractions of objects in the infrared AGN
wedge if the \buxs and SDSS QSO2 surveys target AGN with different
host galaxy properties, in the sense that the infrared emission in the
\buxs type 2 AGN is less contaminated by the host galaxy than in the
SDSS QSO2s at the same luminosity. We will come back to this point in
Sec.~\ref{bias}. The results presented in this section indicate
however, that at high luminosities the infrared AGN wedge seems to be
able to unbiasedly select both hard X-ray type 2 AGN and SDSS QSO2s.

\subsection{Can MIR AGN selection techniques identify luminous heavily obscured AGN?}
\label{compton-thick}
The results in Sec.~\ref{comparison} suggest that at luminosities
${\rm L_{2-10\,keV}>10^{44}\lum}$, the infrared AGN wedge is highly
effective at selecting obscured AGN. However, the sample of \buxs AGN
does not include any Compton-thick objects. To investigate this we
have evaluated whether the SDSS QSO2s identified as Compton-thick
candidates in the literature from the studies of \citet{vignali10} and
\citet{jia12} would be selected by the infrared AGN wedge. The
Compton-thick classification for these objects is based on their low
X-ray/\oiii\, flux ratios (observed 2-10 keV rest-frame luminosities
$\lesssim$1\% of the predicted values based on the
\oiii\,$\lambda$5007 line luminosity).

The X-ray follow-up of the SDSS QSO2s in R08 focused mainly on the
most luminous objects. Hence, to include as many objects as possible
with available X-ray information, we selected all SDSS QSO2s with
\wise detections with \sn$\geq$5 at all 3.4, 4.6, and 12\,\nmic{} and
\loiii$\geq$4.8$\times$$10^{42}\,\lum$ (136 objects in total). The
fraction of those objects in the infrared AGN wedge is ${\rm
  72.8_{-6.5}^{+5.9}}\%$ (99 objects). In total 31 out of 136 objects
have observations in the \xmm and \chandra archives and 24 (${\rm
  77.4_{-13.4}^{+10.4}}\%$) fall in the infrared AGN wedge. This
fraction is entirely consistent with that derived for all the objects
at the same luminosity showing that by selecting the SDSS QSO2s with
X-ray follow-up we are not introducing any bias in terms of infrared
colours.

\begin{figure*}
  \centering
  \begin{tabular}{cc}
    \hspace{-0.7cm}\includegraphics[angle=90,width=0.49\textwidth]{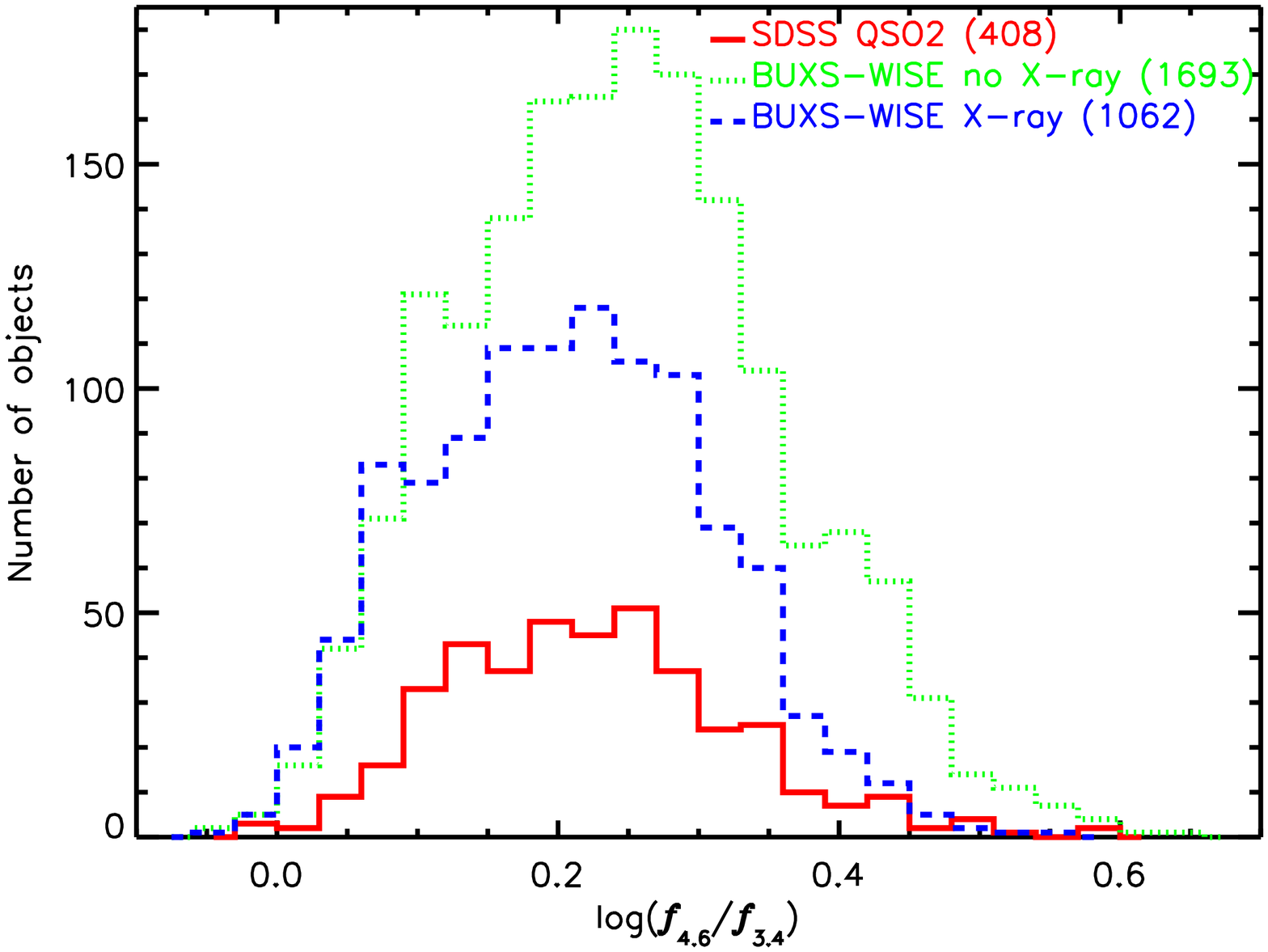}
    \hspace{-0.7cm}\includegraphics[angle=90,width=0.49\textwidth]{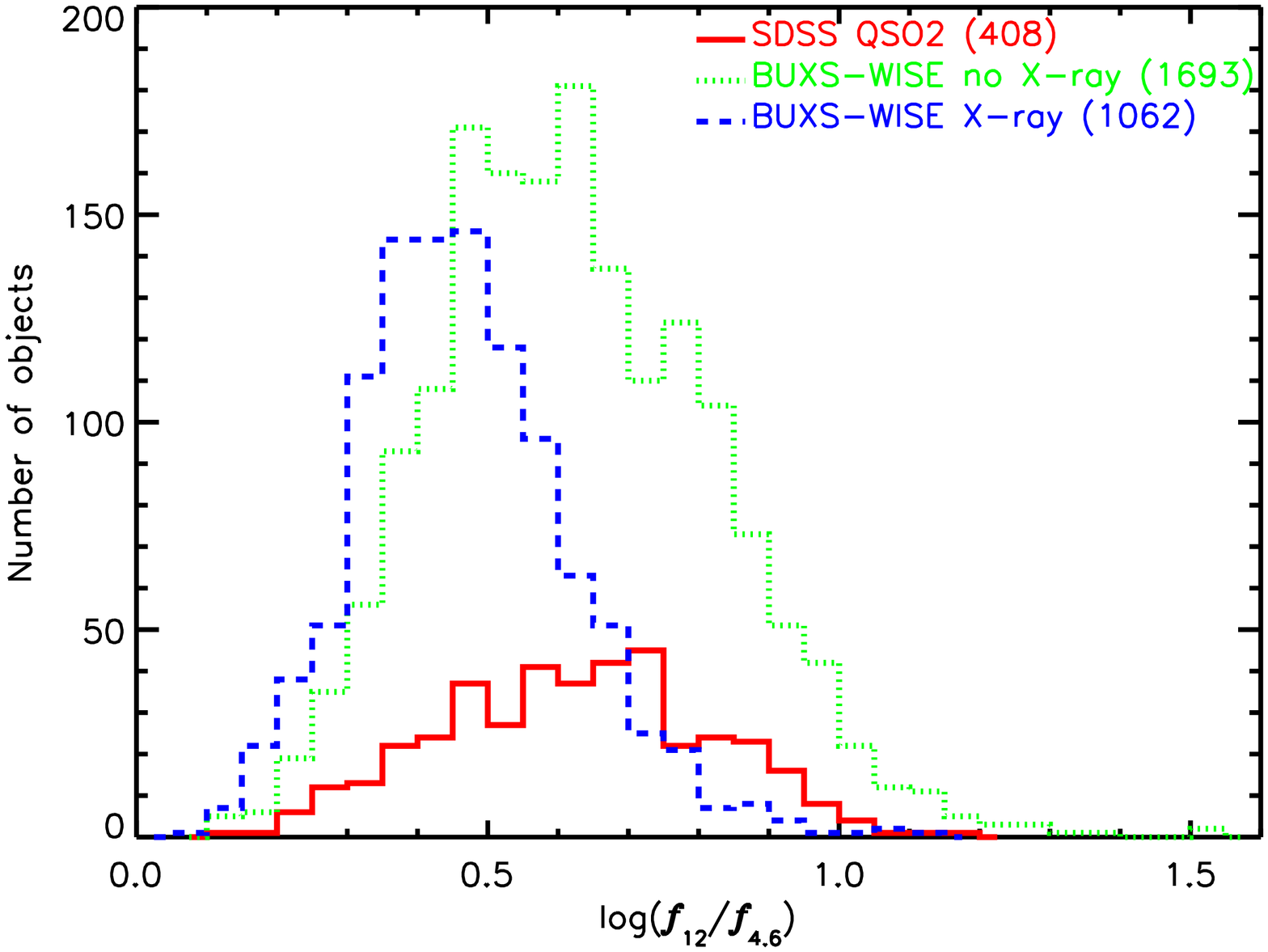}\\
    \hspace{-0.7cm}\includegraphics[angle=90,width=0.49\textwidth]{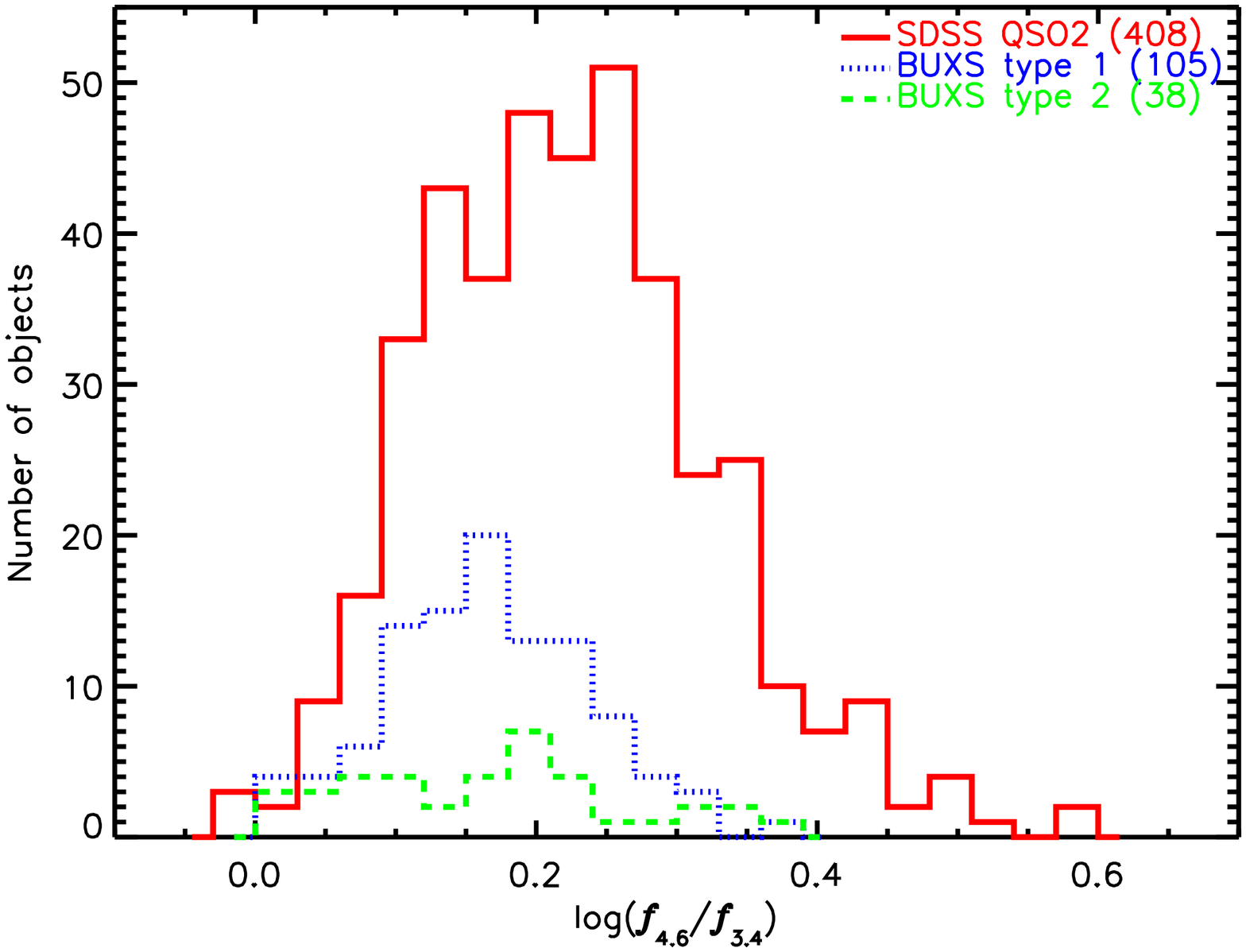}
    \hspace{-0.7cm}\includegraphics[angle=90,width=0.49\textwidth]{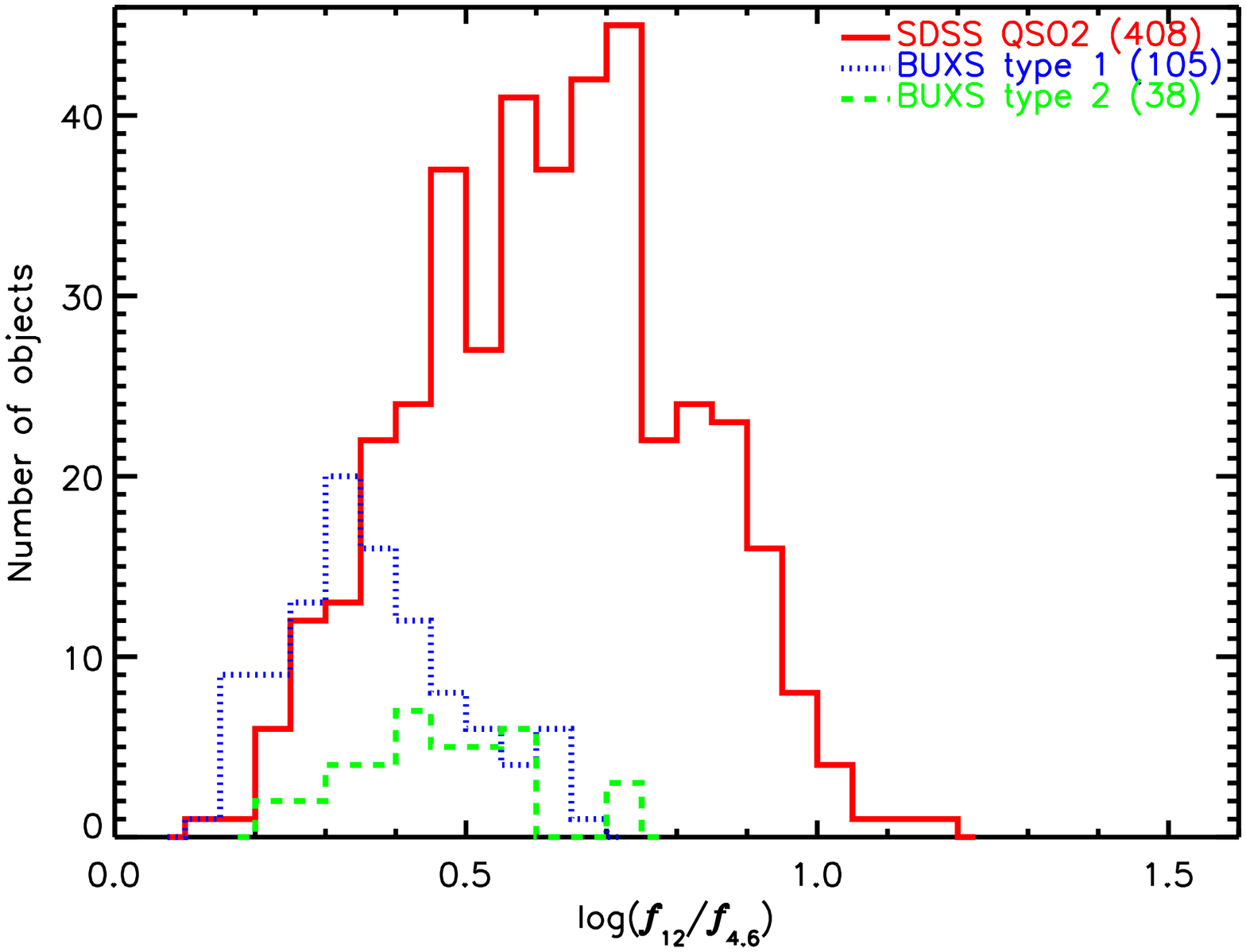}\\
  \end{tabular}
\vspace{+0.05in}
  \caption{Distributions of \wise colours for different classes of
    objects in the infrared AGN wedge. Top: \colY{} (left) and \colX{}
    (right) colour distributions for the SDSS QSO2s (solid line
    histograms) and for all the \wise objects in the \buxs area
      (i.e. not applying any of the X-ray selection criteria that
      define the \buxs AGN sample, see Sec.~\ref{wise_buxs} and
      Table~\ref{tab0}) with and without detection in X-rays,
      respectively (dashed and dotted line histograms). Bottom:
    \colY{} (left) and \colX{} (right) colour distributions for the
    SDSS QSO2s (solid line histograms) and for the type 1 and type 2
    AGN in the \buxs survey (dotted and dashed line histograms).}
  \label{fig5}
\end{figure*}

Out of the 31 SDSS QSO2s with X-ray follow-up, 18 objects are
identified as Compton-thick candidates. Of these 12 are in the
infrared AGN wedge (${\rm 66.7_{-18.5}^{+15.5}}\%$). Although the
number of objects in the comparison is small, the fraction of
Compton-thick QSO2 candidates in the infrared AGN wedge is consistent,
within the uncertainties, with the value for the full SDSS QSO2 sample
at the same luminosities (both with and without X-ray
observations). This suggests that at least at the AGN luminosities
involved in the comparison, Compton-thick and Compton-thin SDSS QSO2s
have similar \wise colour distributions. To illustrate this result, we
show in Fig.~\ref{fig4b} the \wise colours of the SDSS QSO2s with
X-ray follow-up and indicate those objects identified as Compton-thick
candidates either from the studies of \citet{vignali10} or
\citet{jia12}. The contours in Fig.~\ref{fig4b} represent the
distribution of colours for all the SDSS QSO2s with
\loiii$\geq$4.8$\times$$10^{42}\,\lum$. As already indicated, the
Compton-thick candidates have a distribution of colours that is
consistent with that for the SDSS QSO2s with same luminosities.

These results fully support that at high AGN luminosities and
\red$\lesssim$1 our MIR selection technique is very effective at
identifying both Compton-thin and Compton-thick AGN.

\section{Properties of X-ray and optically selected QSO2}
\subsection{Infrared colours} \label{colours}
As shown in Fig.~\ref{fig3}, the \wise objects in the \buxs area
detected in hard X-rays and the SDSS QSO2s preferentially occupy
different regions of the infrared colour-colour diagram. SDSS QSO2s
have overall bluer \colY{} colours and redder \colX{} colours than the
X-ray detected \wise objects. To study this effect in more detail in
what follows we restrict our analysis to the objects in the infrared
AGN wedge. In these objects the MIR fluxes measured with \wise should
be dominated by the AGN.

Fig.~\ref{fig5} (top) shows the distributions of the \colY{} and
\colX{} colours for the SDSS QSO2s and for all the \wise sources in
the \buxs survey area in the infrared AGN wedge with and without
detection in X-rays, respectively. In the infrared AGN wedge X-rays
preferentially trace the objects with the bluest \colY{} and \colX{}
colours. Some caution must be taken however, when using \colY{}
colours as even for objects in the infrared AGN wedge, starlight could
contaminate the \wise measured fluxes at the shortest
wavelengths. Thus in the following we restrict the comparison to
\colX{} colours that should provide a ``cleaner" view of the AGN
component.

A Kolmogorov-Smirnov (K-S) test indicates that the \colX{} colour
distributions of the \wise X-ray non-detected and the SDSS QSO2s are
consistent with each other at a $>$99\% confidence level (null
hypothesis probability 73\%). The SDSS QSO2s in R08 trace better than
X-ray selection the X-ray-undetected infrared AGN candidates with the
reddest colours in the infrared AGN wedge.

Fig.~\ref{fig5} (bottom) shows the \colX{} (and \colY{} for
comparison) colour distributions for the SDSS QSO2s and for the \buxs
type 1 and type 2 AGN, respectively. The \buxs type 2 AGN have overall
redder \colX{} colours than the \buxs type 1 AGN, although the
dispersion of the distributions is large ($\sigma$$\sim$0.12). The
mean colour $\langle$\colX$\rangle$ is 0.37 and 0.43 for the type 1
and type 2 AGN, respectively. A K-S test between the type 1 and type 2
AGN colour distributions shows that they are different at a $>$99\%
confidence level (null hypothesis probability 0.13\%). We obtain the
same result when comparing the \colX{} colour distributions of the
\buxs type 2 AGN and the SDSS QSO2s. The SDSS QSO2s trace the objects
with the reddest \colX{} colours (and also reddest \colY{} colours) in
the infrared AGN wedge, with $\langle$\colX{}$\rangle$=0.62 and
$\sigma$$\sim$0.19.

\subsubsection{Redshift and luminosity effects} \label{z_l}
\begin{figure}
  \centering
  \begin{tabular}{cc}
    \hspace{-0.7cm}\includegraphics[angle=90,width=0.49\textwidth]{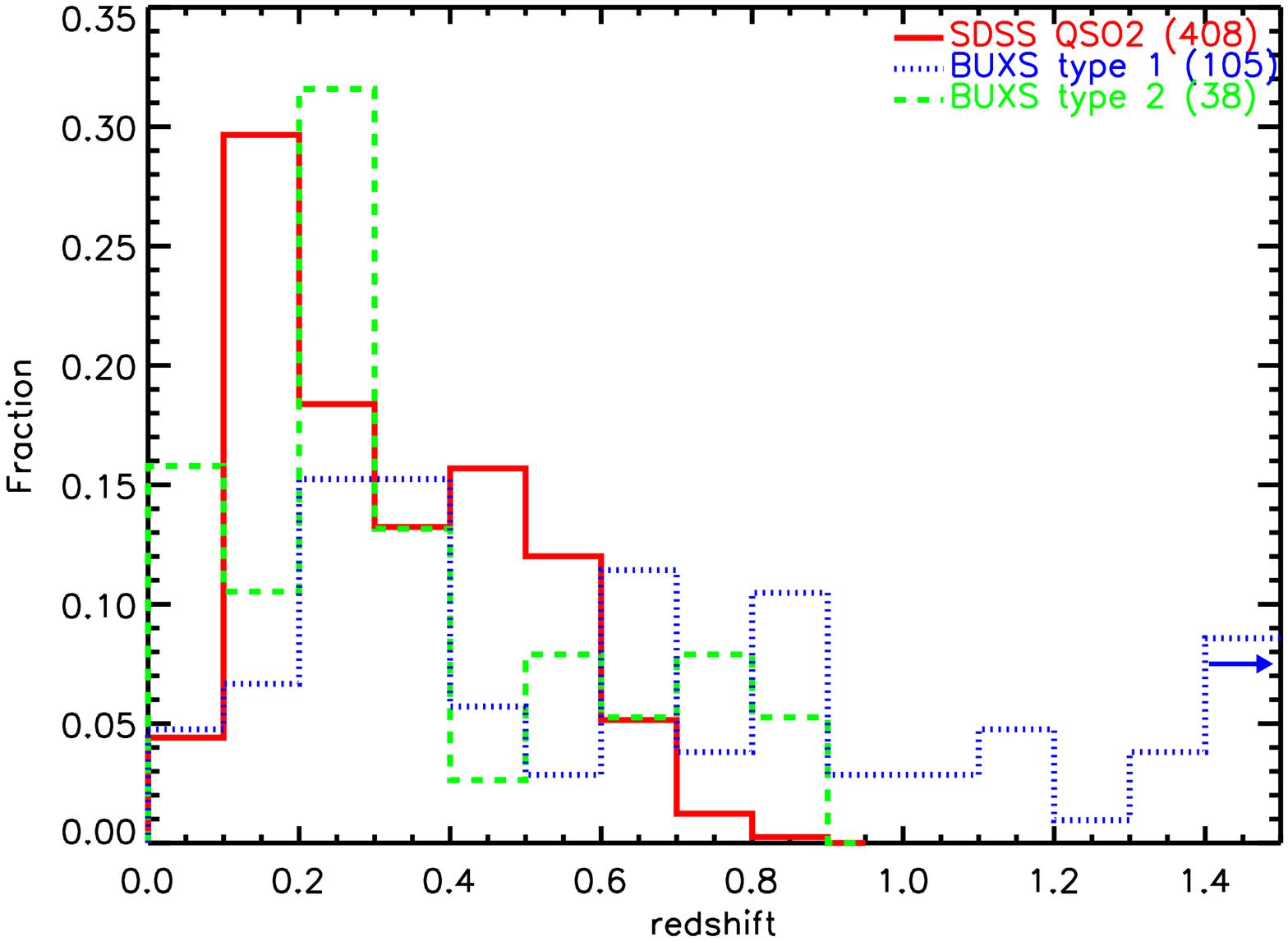}\\
    \hspace{-0.7cm}\includegraphics[angle=90,width=0.49\textwidth]{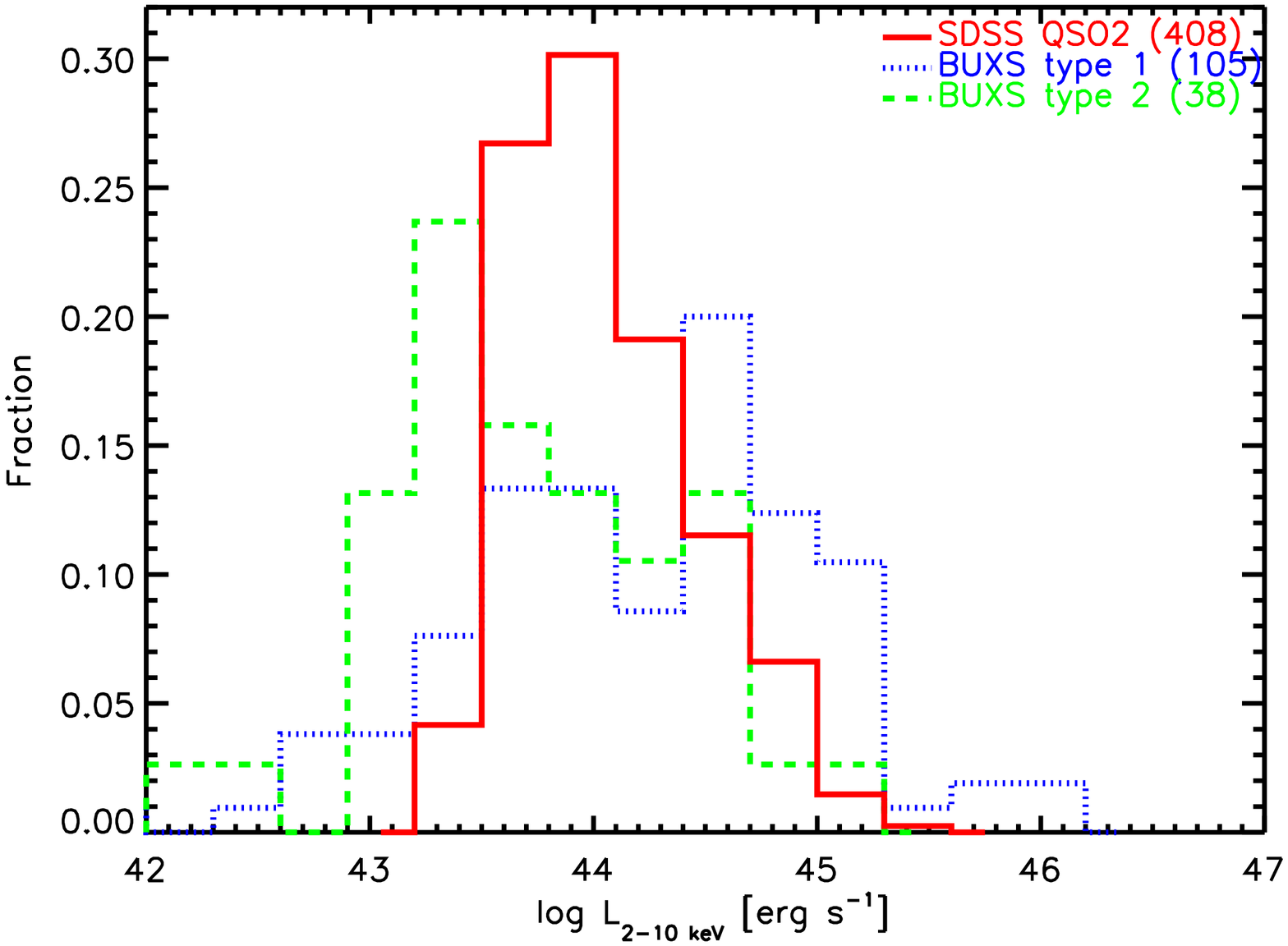}\\
    \hspace{-0.7cm}\includegraphics[angle=90,width=0.49\textwidth]{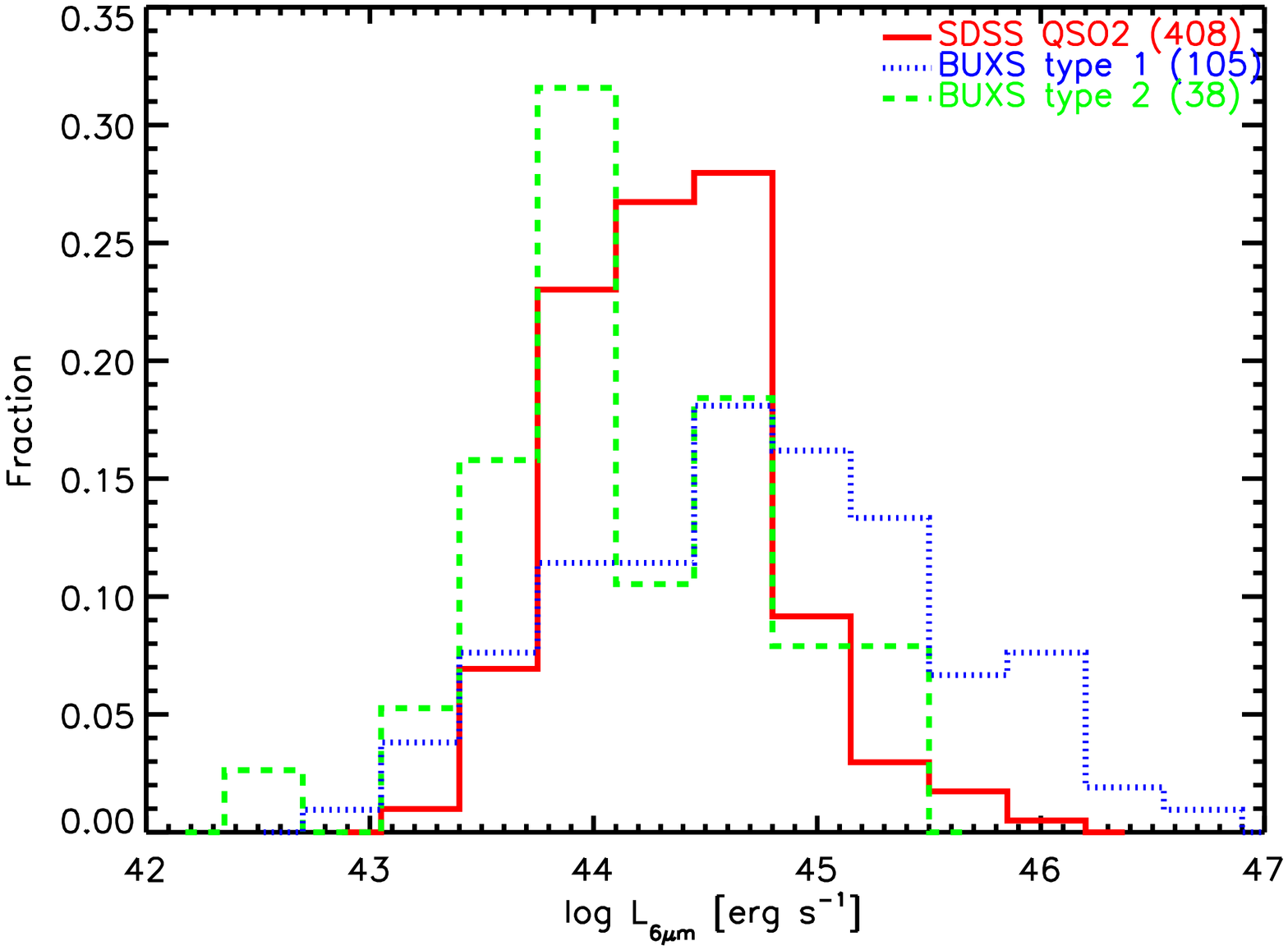}\\
  \end{tabular}
\vspace{+0.05in}
  \caption{Comparison of the \red{} (top), 2-10 keV (middle) and
    6\,\mic luminosity (in logarithmic units) distributions for the
    SDSS QSO2s in R08 and the \buxs type 1 and type 2 AGN,
    respectively. For the SDSS QSO2s the 2-10 keV luminosities were
    derived from the empirical relation between X-ray and
    \oiii\,$\lambda$5007 luminosities of J12 (see
    Sec.~\ref{comparison}). For the \buxs AGN the X-ray luminosities
    were derived from the modeling of their X-ray spectra and are
    corrected for intrinsic absorption. The monochromatic 6\,\mic
    (rest-frame) luminosities were derived using linear interpolation
    between \wise nearby bands. }
  \label{fig6}
\end{figure}

In order to compare the observed infrared colours of objects it is
important that they span the same range of \red{} and
luminosities. This way we minimize k-correction effects and we account
for the luminosity dependence of host galaxy contamination to the
observed MIR fluxes (e.g. \citealt{cardamone08}; \citealt{assef12};
\citealt{donley12}; M12; \citealt{stern12}). Fig.~\ref{fig6} shows the
\red{} (top) and 2-10 keV luminosity (middle) distributions of the
\buxs AGN and the SDSS QSO2s. The \buxs type 2 AGN and SDSS QSO2s span
the same \red\, range, therefore it does not appear that k-correction
effects alone can explain their different MIR colour distributions.

On the other hand, according to the predicted 2-10 keV
luminosities\footnote{We have not used in the comparison \oiii\,
  luminosities for the \buxs AGN as not all the spectra sample the
  rest-frame wavelengths of the \oiii\, line and in many cases the
  \sn\, is not sufficient to obtain accurate measurements of the line
  luminosity.} the SDSS QSO2s are intrinsically more luminous than the
\buxs type 2 AGN. A K-S test gives a probability higher than
  99.99\% than the two distributions are different. This difference
should be somewhat larger if we consider that the \oiii\,$\lambda$5007
luminosities were not corrected for extinction (see
Sec.~\ref{comparison}). To confirm this result in Fig.~\ref{fig6}
(bottom) we show the distributions of rest-frame 6\,\mic monochromatic
luminosities. At these wavelengths the emission should suffer little
contamination from star formation processes in the host galaxy
(e.g. \citealt{nardini10}) and should be much less affected by dust
extinction than the optical \oiii{} line. Therefore, rest-frame
luminosities at 6\,\mic are a robust independent proxy of the AGN
intrinsic power. The 6\,\mic luminosities were computed from linear
interpolation of the \wise observed fluxes (see Sec.~\ref{seds} for
more details). A K-S test gives a 99.5\% probability that the 6\,\mic
luminosity distributions for the \buxs type 2 AGN and the SDSS QSO2s
are different. 

Both monochromatic 6\,\mic (rest-frame) continuum and 2-10 keV
predicted luminosities indicate that the SDSS QSO2s are on average
$\sim$20-30\% more luminous than the BUXS type 2 AGN. Although the
overlap in the luminosity distributions is large, it is reasonable to
expect that at the shortest \wise wavelengths starlight contamination
from the blue infrared hosts might be more important in the
intrinsically less luminous \buxs type 2 AGN.

The type 1 AGN in \buxs have the highest luminosities in this
comparison and the bluest infrared colours. Because the type 1 AGN
span a broad range of \red, their blue MIR colours could be the result
of sampling shorter rest-frame wavelengths. To evaluate this effect we
have compared the \colX{} colour distributions for the type 1 AGN at
\red$\leq$0.6 and \red$>$0.6 (similar number of objects on each
bin). A K-S test indicates that the two colour distributions are
consistent (null hypothesis probability 13.5\%). Higher \red{} (more
luminous) type 1 AGN tend to have redder infrared colours. Thus,
k-correction effects cannot explain the different colour distributions
of the BUXS type 1 AGN and the X-ray and optically selected type 2 AGN
in the infrared AGN wedge.

\subsection{Median spectral energy distributions} \label{seds}
To understand the nature and properties of the AGN selected using
\wise colours, we have computed the median UV-to-MIR rest-frame SEDs
of the \buxs AGN and SDSS QSO2s in the infrared AGN wedge. Ideally
this should be performed on the individual SEDs, to avoid the
uncertainties associated with the important dispersion in SED
properties even amongst objects in the same AGN class
(e.g. \citealt{krawczyk13}; \citealt{elvis12}; \citealt{lusso11};
\citealt{richards06}). Unfortunately, with the existing data we do not
have sufficient spectral coverage, especially at infrared wavelengths
(only four \wise points from 3 to 22\,\nmic), to properly characterize
the individual SEDs.

\begin{figure*}
  \centering
  \begin{tabular}{cc}
    \hspace{-1cm}\includegraphics[angle=90,width=0.6\textwidth]{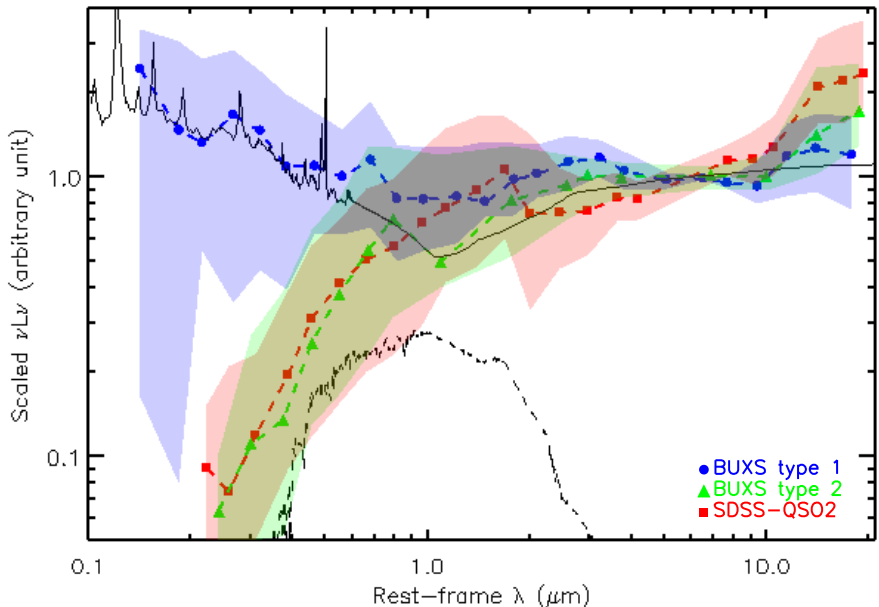}
  \end{tabular}
  \caption{Median UV-to-MIR SEDs of the different classes of AGN in
    the infrared AGN wedge. The SEDs are normalised at 6\,\mic
      rest-frame. Squares for the SDSS QSO2s, circles for the \buxs
      type 1 AGN and triangles for the \buxs type 2 AGN,
      respectively. The points are connected with a power-law in
      linear space (dashed lines). The shaded regions in the figure
      represent the dispersion in the data points quantified with the
      68.27\% percentiles. For comparison we show two templates from
      the library of \citet{polletta07}: a type 1 QSO normalised at
      6\,\mic rest-frame (solid line) and an elliptical galaxy with an
      arbitrary normalization (dashed line).}
  \label{fig7}
\end{figure*}

For each object we make use of the \wise all-sky data release in
combination with optical photometry from SDSS in the $u$, $g$, $r$,
$i$ and $z$ bands (\citealt{fukugita96}) and observations in the {\it
  Y}, {\it J}, {\it H} and $K$ near-infrared bands from the UKIDSS
Large Area Survey (LAS) ninth data release (\citealt{lawrence07}).
For details on the UKIDSS photometric system, calibration and pipeline
processing please refer to \citet{hewett06}, \citet{hodgkin09} and
\citet{hambly08}, respectively. A conservative 5\% error was added in
quadrature to the SDSS and UKIDSS catalogued flux errors to account
for the uncertainties in the zero-points (for \wise fluxes see
Sec.\,\ref{wise_data}). To obtain the near-infrared counterparts of
the SDSS QSO2s we cross-matched the optical positions with the UKIDSS
LAS using a 2 arcsec search radius. UKIDSS data is available for
$\sim$41\% of the SDSS QSO2s. We note that $\sim$54\% of the SDSS QSO2s
have detections in at least one of the Two-Micron All-Sky Survey
(2MASS; \citealt{skrutskie06}) filters, most of these objects being at
\red$<$0.3. However, to avoid biasing the median SEDs at rest-frame
near-infrared wavelengths towards the objects with the brightest host
galaxies (i.e. near-infrared bright SED shapes), we only used the data
from UKIDSS that it is three magnitudes deeper than 2MASS.

To compute the median SEDs we shifted the individual SEDs to a common
rest-frame wavelength grid. Because the aim of this analysis is to
investigate the overal shape of the observed SEDs, we do not apply any
corrections for host galaxy contamination. However the fluxes are
corrected for Galactic reddening.  We converted the fluxes to
luminosities in ${\rm \nu L_\nu}$ units and finally, we normalised the
SEDs at 6\,\mic rest-frame. At this wavelength we should get a
relatively uncontaminated view of the AGN while the effects of
extinction should be small (e.g. \citealt{lutz04}). The 6\,\mic
luminosities were computed with linear interpolation between nearby
points in log{}${\rm \nu L_\nu}$ versus log{}${\rm \lambda}$
space. The data were combined and grouped in wavelength bins
containing at least 15 points. We required a minimum bin size of
0.08\,\mic in logarithmic wavelength. Linear interpolation was used to
convert the photometry to the grid points. The luminosity on each bin
is the median of the data points, while the dispersion is represented
with the percentiles for a 68.27\% confidence.

\begin{figure}
  \centering
  \begin{tabular}{cc}
    \hspace{-2.3cm}\vspace{1.8cm}\includegraphics[angle=-90,width=0.36\textwidth]{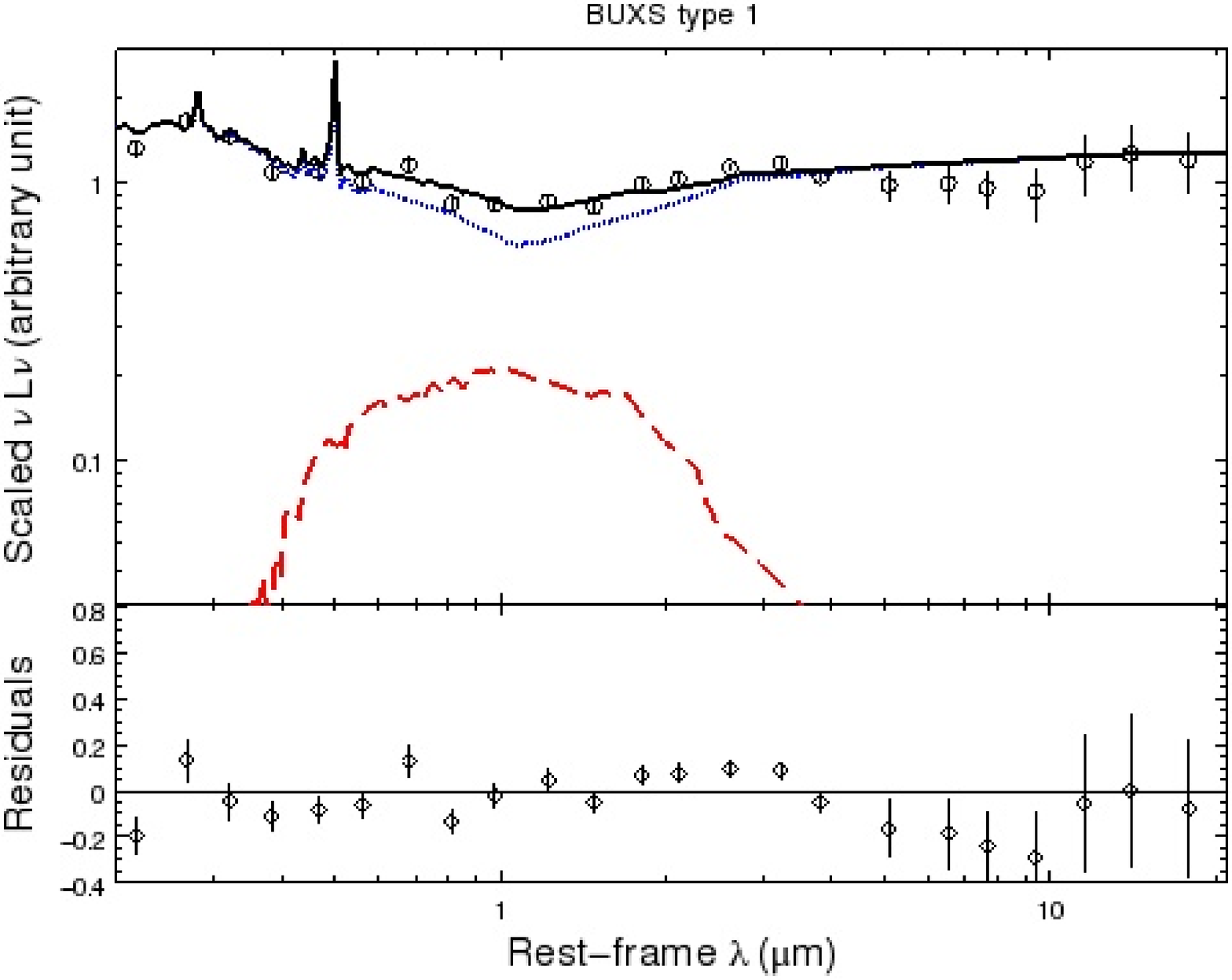}\\
    \hspace{-2.3cm}\vspace{1.8cm}\includegraphics[angle=-90,width=0.36\textwidth]{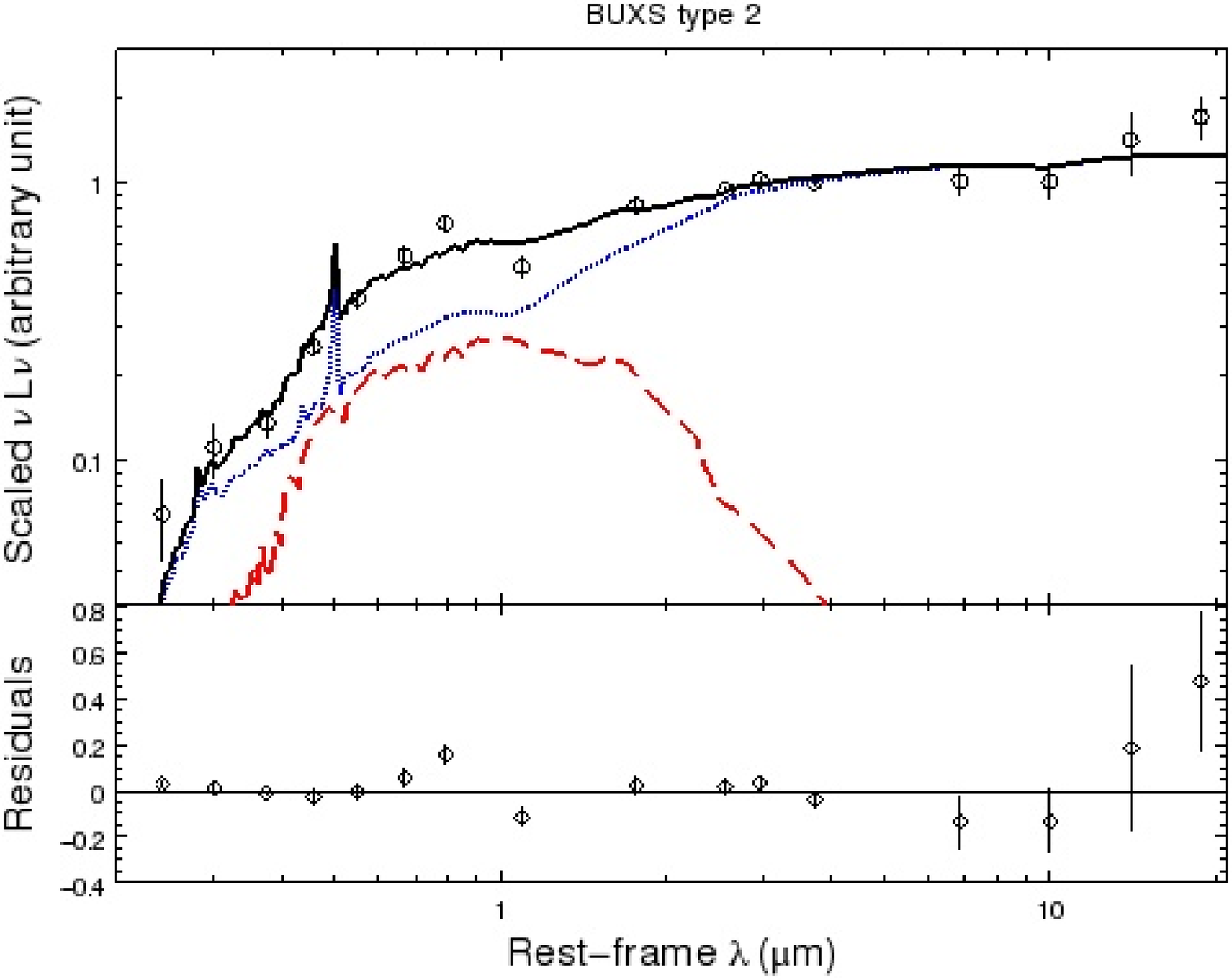}\\
    \hspace{-2.3cm}\includegraphics[angle=-90,width=0.36\textwidth]{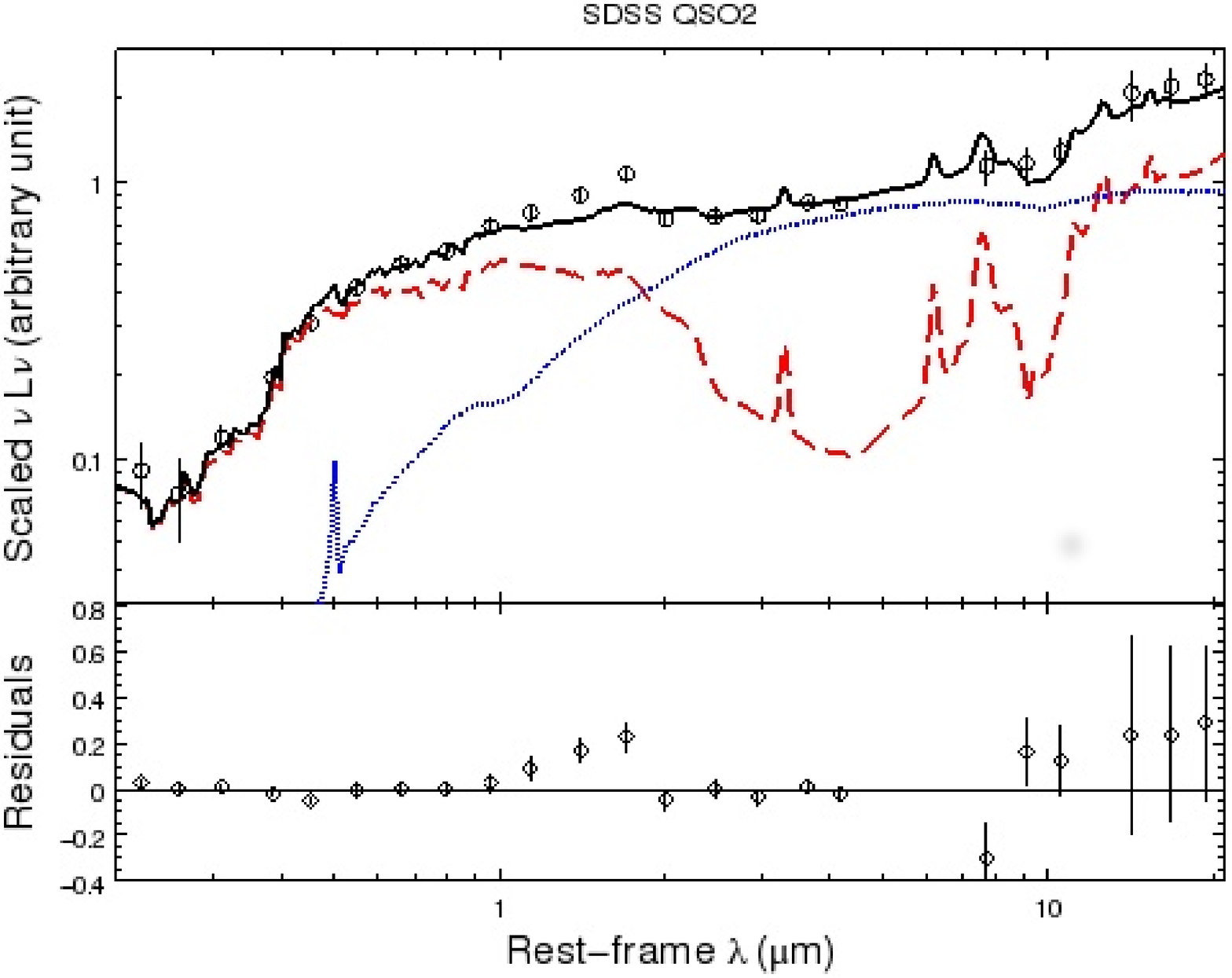}\\
  \end{tabular}
\vspace{+0.8in}
  \caption{Median SEDs of the \buxs type 1 (top) and type 2 (middle)
    AGN and the SDSS QSO2s (bottom) that fall in the infrared AGN
    wedge. The data points are indicated with open circles.  The error
    bars are the mean of the flux errors for the individual data
    points in the bin. The best-fit models are illustrated with the
    solid lines while the AGN and host galaxy components are
    illustrated with the dotted and dashed lines, respectively.}
  \label{fig8}
\end{figure}

Fig.~\ref{fig7} shows the median SEDs in ${\rm \nu L_\nu}$ versus
$\lambda$. For comparison we show a type 1 AGN template SED and an
elliptical galaxy from the Spitzer Wide-area InfraRed Extragalactic
survey (SWIRE) library of \citet{polletta07}. At rest-frame
UV/optical wavelengths the \buxs type 1 AGN have the characteristic
blue bump continuum typical of the primary emission from the central
accretion disk with a change in slope at around $\sim$1-2\,\nmic{}
rest-frame. At these wavelengths, however, the median SED is rather
flat not showing a clear $\sim$1\,\mic inflection point between the UV
and near-infrared bumps (\citealt{elvis12};
\citealt{richards06}). This is most likely due to significant
contamination from the hosts. This is not surprising as host galaxy
corrections can be important even for luminous type 1 AGN
(e.g. \citealt{elvis94}). For the \buxs type 2 AGN and SDSS QSO2s
their red UV/optical continuum emission is most likely due to
starlight and/or a reddened AGN. This is expected as the
classification as a type 2 AGN is based on the lack of sight of the
nuclear emission and broad-line region. Fig.~\ref{fig7} shows that the
near-infrared stellar bump at 1.6\mic is clearly more prominent in the
SDSS QSO2s. As the median SEDs are normalized at 6\,\mic rest-frame
luminosities where the emission should be dominated by the AGN, this
suggests that the SDSS QSO2s in R08 reside in more luminous hosts.

 At wavelengths $\gtrsim$2\,\mic (rest-frame) the median SED of the
 type 1 AGN shows an apparent infrared-bump centered at around
 3\,\nmic. This feature has been interpreted as the signature of the
 hottest dust component originating in the inner parts of the
 obscuring torus (see e.g. \citealt{neugebauer79};
 \citealt{glikman06}; \citealt{netzer07}). At these wavelengths the
 median SED of the \buxs type 2 AGN is marginally redder than that of
 the \buxs type 1 AGN, and the SDSS QSO2s have the reddest
 continuum. These results are entirely consistent with the observed
 differences in the distributions of their infrared colours (see
 Sec.~\ref{colours} and Fig.~\ref{fig5}).
 
 Above $\sim$10\,\mic all the SEDs show an apparent excess of MIR
 emission when compared with the type 1 AGN template. This excess is
 more significant for the SDSS QSO2s. Although the \wise survey at
 22\,\mic is significantly shallower than those at 3.4, 4.6, and
 12\,\nmic, most SDSS QSO2s and \buxs AGN have 22\,\mic detections with
 at least \sn$\geq$2\footnote{Below \sn=2 \wise flux measurements are
   considered upper limits \citep{cutri12}.}. For the SDSS QSO2s,
 93.4\% have 22\,\mic detections with \sn$\geq$5 and all but one have
 22\,\mic detections with \sn$\geq$2. Thus, it is unlikely that the
 median SEDs are biased towards the brightest objects at 22\,\nmic.

\subsubsection{SED fitting with templates} \label{fit}
 To provide a clearer description of the median SEDs we fitted the
 data with a combination of the type 1 AGN (QSO1) template and one of
 the galaxy templates from the library of \citet{polletta07}. The
 galaxy templates we used comprise elliptical, spiral as well as the
 starburst templates for I22491, M82, N6090 and Arp220. We selected
 the fit with the lowest reduced $\chi^2$. To account for obscuration
 of the nuclear AGN light we applied the Small Magellanic Cloud
 (SMC)-like dust-reddening laws from \citet{gordon03} below 8100\AA{}
 and \citet{cardelli89} at longer wavelengths. With this approach we
 allow two distinct components, a dusty central structure (warm
 AGN-heated dust) and an extended star-forming region (cold
 starburst-heated dust), to contribute to the infrared emission, and a
 screen-like dusty absorber that reddens the AGN nuclear emission.
 
In a more physically motivated characterisation of the SEDs the
UV/optical emission from the accretion disk and the infrared emission
from the torus should be modelled independently. A screen-like nuclear
extinction should only be applied to the accretion disk emission while
the torus parameters (e.g. viewing angle, optical depth, number of
clouds, opening angle, torus size) are incorporated in the different
AGN torus model realizations. Unfortunatelly, these models, especially
those where the absorber is clumpy which are strongly supported by
observations, have a very large number of parameters
(e.g. \citealt{nenkova02}; \citealt{honig06}; \citealt{nenkova08a};
\citealt{schartmann08}; \citealt{heymann12}). Our median SEDs do not
have sufficient spectral resolution to obtain any meaningful
constraints on the parameters of the dusty torus. However, as we will
see in Sec.~\ref{x_seds} and Sec.~\ref{xo_seds}, the best-fit
extinctions are small and therefore it is reasonable to assume that
the screen-like absorber in our models only has an effect on the
accretion disk emission component. As the goal of this analysis is to
obtain a rough description of the median SEDs by disentangling the AGN
and host galaxy emission, the best-fit parameters should only be used
as a guide of the overall behaviour in the different classes of
objects. The results of the SED fitting are shown in Fig.~\ref{fig8}.

\subsubsection{SEDs of X-ray selected type 1 and type 2 AGN} \label{x_seds}
 The QSO1 template from \citet{polletta07} provides a fair
 representation of the median UV to MIR SED of the \buxs type 1 AGN
 (best-fit ${\rm \chi^2_\nu}$=3.4). At rest-frame
 optical/near-infrared wavelengths there is contamination from
 starlight. This component is best modelled with an elliptical galaxy
 template. At MIR wavelengths the emission is clearly dominated by the
 AGN although the residuals of the fit suggest that the continuum is
 somewhat bluer than in the QSO1 template.

For the \buxs type 2 AGN both stellar emission (also best modelled
with an elliptical galaxy template) and a reddened AGN continuum
(best-fit ${\rm N_H=7.9\pm0.4\times10^{21}\,cm^{-2}}$ or ${\rm
  A_V}$$\sim$0.6 assuming the SMC dust-to-gas ratio;
\citealt{gordon03}) are required to provide a good quality fit (${\rm
  \chi^2_\nu}$=2.4) at rest-frame UV/optical wavelengths. The MIR
emission is dominated by the AGN and there is no evidence of
star-forming activity. This is an expected result as the median SEDs
were derived using only the objects in the infrared AGN wedge.

As a test we also fitted the \buxs type 2 AGN median SED with the QSO2
SED template from \citet{polletta07} as it should provide a more
accurate representation of the typical MIR continuum for luminous type
2 AGN. This template also provides a good quality fit at MIR
wavelengths and confirms that the MIR continuum of the \buxs type 2
AGN in the infrared AGN wedge is dominated by the AGN component.

\begin{figure*}
  \centering
  \begin{tabular}{cc}
    \hspace{-0.5cm}\includegraphics[angle=90,width=0.42\textwidth]{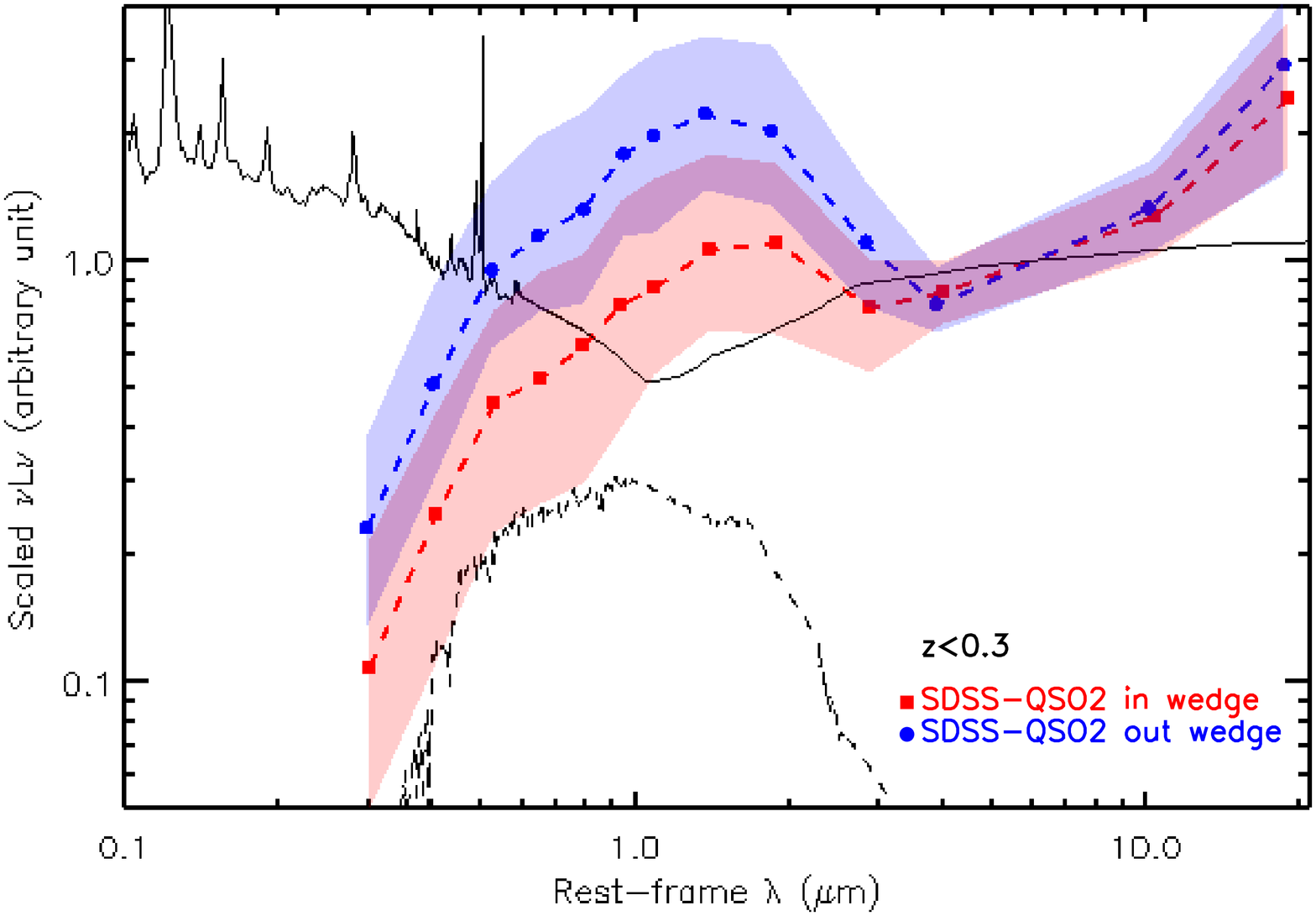}
    \hspace{+1.2cm}\includegraphics[angle=90,width=0.42\textwidth]{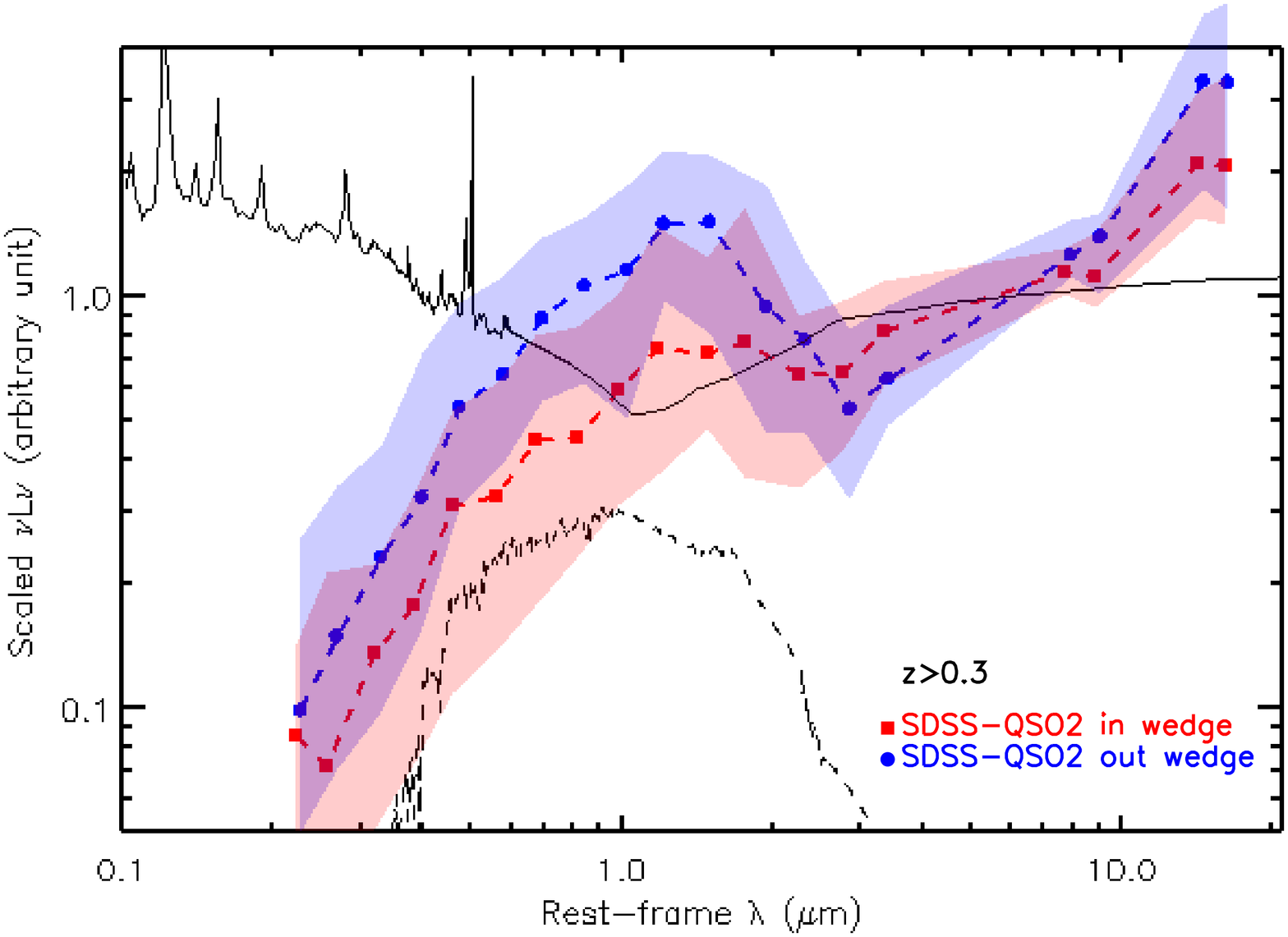}
  \end{tabular}
  \caption{Median UV-to-MIR SEDs for the SDSS QSO2s in and out the
    infrared AGN wedge, respectively. The SEDs are normalised at
      6\,\mic rest-frame. Left: SDSS QSO2s at \red$\leq$0.3. Right:
    SDSS QSO2s at \red$>$0.3. The points are connected with a
    power-law in linear space (dashed lines). The shaded regions in
    the figure represent the dispersion in the data points quantified
    with the 68.27\% percentiles. For comparison we show two
      templates from the library of Polletta et al. (2007): a type 1
      QSO normalised at 6\,\mic rest-frame (solid line) and an
    elliptical galaxy with an arbitrary normalization (dashed line).}
  \label{fig11}
\end{figure*}

\subsubsection{SEDs of SDSS QSO2s} \label{xo_seds}
At the wavelengths sampled by our data the contribution from the host
galaxy to the median SED of the SDSS QSO2s is more important than in
the \buxs type 2 AGN. An acceptable fit could only be obtained with a
composite SED where a reddened AGN continuum (best-fit ${\rm
  N_H=1.3\pm0.2\times10^{22}\,cm^{-2}}$ or ${\rm A_V}$$\sim$1 assuming
the SMC dust-to-gas ratio; \citealt{gordon03}), stellar emission and
features associated with star formation activity contribute to the
observed fluxes. The best-fit was obtained with the star-forming
galaxy SED of M82 (${\rm \chi^2_\nu}$=2.2 ; see Fig.~\ref{fig8},
bottom). Still the MIR emission in these objects is dominated by the
AGN. This is expected as the median SED was derived for the SDSS QSO2s
in the infrared AGN wedge.

At optical/near-infrared wavelengths both the AGN and stellar emission
are required to provide a good quality fit although the AGN
contribution is not as important as in the \buxs type 2 AGN. None of
the star-forming templates in the library of \citet{polletta07} can
fit the near-infrared stellar bump. The most plausible explanation for
this result is that we are combining the SEDs of a large number of
objects with a range in host galaxy properties. We also checked that
the QSO2 template from \citet{polletta07} also provides a good quality
fit to the data points at MIR wavelengths, reaching similar
conclusions.

In Sec.~\ref{colours}, from the comparison of the \colX{} colour
distributions of the \buxs type 1 and type 2 AGN and the SDSS QSO2s,
we found a trend where higher (line of sight) absorption might be
associated with redder infrared colours. However we have shown that,
even for luminous AGN in the infrared AGN wedge, host galaxy
contamination of the \wise fluxes can have an effect on the observed
properties of the objects. Our results suggest that it is not the
possible presence of Compton-thick AGN that determines the red MIR
continuum of the SDSS QSO2s but most likely the properties of their
host galaxies. 

\section{Discussion} \label{discussion}
The results of the analysis of the median SEDs of the SDSS QSO2s are
entirely consistent with those obtained from \spitzer MIR spectroscopy
of QSO2s selected from optical SDSS observations
(e.g. \citealt{zakamska08}). The MIR spectra of these objects show a
large diversity of properties. While some objects show a featureless
continuum, star formation activity is detected in most QSO2s although
the MIR emission is dominated by the AGN. In the study of a sample of
SDSS galaxies at \red$<$0.3 \citet{kauffmann03} suggested a connection
between the AGN power and star formation activity. A young stellar
population and signatures of recent bursts of star formation are a
general property for the most powerfull QSO2s. A similar result was
found by \citet{vandenberk06} for SDSS selected type 1 AGN at
\red$<$0.75. This is consistent with our finding that star formation
is a general property in the most luminous (the sources that
preferentially enter the infrared AGN wedge) SDSS QSO2s.

\subsection{Understanding the MIR continuum emission of SDSS QSO2s} \label{bias}
As noted in Sec.~\ref{seds} (see also Fig.~\ref{fig7} and
Fig.~\ref{fig8}) the SDSS QSO2s have more pronounced near-infrared
stellar bumps. We explain this as a selection effect due to the tight
magnitude limits applied by the SDSS spectroscopic target selection
algorithms. For example, the Galaxy algorithm selects resolved objects
with Petrosian magnitudes $r$$<$17.77 mag while the Low-\red{} and
High-\red{} QSO algorithms target objects with point spread function
(PSF) magnitudes 15$<$$i$$<$19.1 and 15$<$$i$$<$20.2,
respectively. The Serendipity FIRST algorithm selects sources with
fiber magnitudes 14$<$$g,\,r,\,i$$<$20.5. For comparison, the \buxs
AGN span a much broader range of optical magnitudes, 15$<r<$23.5,
where $r$ refers to SDSS PSF magnitudes. Because the host galaxy
stellar emission is an important component at UV/optical wavelengths
in type 2 AGN, the magnitude limits of the SDSS spectroscopy are
likely biasing the samples towards type 2 objects with more luminous
(more massive) hosts. This can reduce the completeness of the infrared
AGN wedge for selection of SDSS QSO2s as such techniques
preferentially miss AGN in massive hosts. This is actually what we
find when comparing the effectiveness of the infrared AGN wedge for
the \buxs type 2 AGN and the SDSS QSO2s (see Sec.~\ref{comparison}):
the selection completeness for the SDSS QSO2s is lower than that of
the BUXS type 2 AGN, at least in the low luminosity regime.

To illustrate the host galaxy dilution effect we show in
Fig.~\ref{fig11} the median SEDs for the SDSS QSO2s at \red$\leq$0.3
(left) and \red$>$0.3 (right) that are in/out the infrared AGN wedge
(normalized at 6\,\mic rest-frame), respectively. The objects outside
the infrared AGN wedge reside in more massive hosts. The host galaxy
emission contaminates the \wise fluxes at the shortest wavelengths, an
effect that becomes more important at higher \red{} due to the
k-correction. Based on the existing data we do not find significant
differences in the MIR continuum of the SDSS QSO2s in and out the
infrared AGN wedge. Although we cannot quantify the role of dust
extinction in shaping the MIR spectra, we would expect that star
formation features would be stronger in the objects with more massive
galaxies. Obscured AGN normally reside in optically luminous massive
galaxies (stellar masses $M_\star\gtrsim10^{10-11}\,M_\odot$;
\citealt{babic07}; \citealt{alonso08}; \citealt{brusa09};
\citealt{mainieri11}). If the SDSS QSO2s follow the correlation
between the specific star formation rate, i.e the star formation rate
per unit stellar mass, and $M_\star$ in local star-forming galaxies
(e.g. \citealt{brinchmann04}; \citealt{salim05}) it is reasonable to
expect the detection of star formation signatures in these objects.

We note that obscuration by gas in the host galaxy on scales of
hundreds of parsecs (e.g. starbursts, edge-on discs) might also be
responsible for the UV/optical reddening measured in the median SEDs
of the \buxs type 2 AGN and the SDSS QSO2s. Observational evidence for
this effect is supported by studies of the host galaxy properties of
optically selected AGN (\citealt{maiolino95}; \citealt{lagos11}). For
example, the study of \citet{lagos11}, based on the SDSS spectroscopic
survey, finds a tendency for type 1 AGN to reside on face-on systems
while type 2 objects, especially those with high \oiii\, equivalent
widths, reside preferentially on edge-on systems. Thus, host galaxy
obscuration might have a more important role in the SDSS QSO2s than in
the \buxs type 2 AGN.

Finally, we stress that it is unlikely that QSO2 samples based on the
\oiii\, line emission, such as the R08 catalogue, are biased towards
objects with more intense star formation activity. As shown in
\citet{kauffmann03}, \oiii\,$\lambda$5007 emission is relatively weak
in metal-rich star-forming galaxies, with star formation contributing
only a few percent to the \oiii\, fluxes. A more recent work by
\citet{lamassa10} finds no strong evidence that \oiii\, selection is
biased towards AGN with strong star formation activity.

\subsection{Identifying AGN with X-ray, optical and MIR surveys}
\label{agn_surveys}
Hard X-ray surveys ($>$a few keV) provide the most complete and
reliable census of AGN since they suffer little contamination from
processes not related to SMBH activity, missing only Compton-thick
sources. As hard X-ray emission is not affected by host galaxy
dilution, these surveys can identify the AGN residing in the most
massive galaxies and/or with more intense star formation and can
identify a large population of low luminosity AGN.

On the other hand, \oiii\,$\lambda$5007 selection is less sensitive to
circumnuclear obscuration. Therefore, it is an efficient and powerful
technique to provide a more complete view of the heavily obscured and
Compton-thick AGN population. However, such selection depends on
empirical emission-line ratio diagnostic diagrams which are known to
missidentify many faint AGN in star-forming galaxies
(\citealt{trouille10}). Indeed, a significant fraction of X-ray
selected AGN, mainly at ${\rm L_{2-10\,keV}<10^{43}}$, have optical
spectra dominated by the host galaxy and without signatures of AGN
activity (e.g. \citealt{caccianiga07}). Thus, \oiii-based surveys
suffer from effects associated with host galaxy dilution. They will be
incomplete at low luminosities and for objects in which the narrow
line region is heavily extincted.

MIR based surveys are a promising technique to identify absorbed AGN
missed in X-ray surveys, as most of the absorbed radiation should be
re-emited by dust at these wavelengths. The results presented in this
paper support that \wise colour-based MIR selection techniques are
able to identify luminous heavily absorbed and
even Compton-thick AGN at least up to \red$\lesssim$1. The best
candidates are those objects in the infrared AGN wedge with the
reddest MIR colours. However, it is important to note that MIR
selection identifies predominantly luminous AGN. At low luminosities,
especially for those objects with large extinction at rest-frame
infrared wavelengths, the selection is largely incomplete and strongly
biased against AGN residing in massive and/or star-forming hosts. This
will be more important at higher \red{} due to the k-correction. This
observational bias needs to be taken into account for example when
studying the coevolution of SMBH and their host galaxies.

\section{Conclusions}
The main goal of this work was to evaluate whether AGN selection
techniques in the infrared with \nwise, such as the infrared AGN wedge
presented in M12, are able to identify with high efficiency absorbed
AGN missed in X-rays. We also studied the properties of the AGN
populations independently selected in X-ray and optical surveys. To
this aim we used the largest catalogue of optically selected QSO2s at
\red$\leq$0.83 in SDSS. These objects were selected on the basis of
their high \oiii\,$\lambda$5007 line luminosities, \loiii${\rm
  \geq10^{8.3} L_\odot}$. Such a selection should be relatively
unbiased against heavy obscuration. We then compared the results with
those for the hard ($>$4.5 keV) X-ray selected AGN in the \buxs
survey.

The fraction of SDSS QSO2s in the infrared AGN wedge is a strong
function of the AGN luminosity. In total, 53.3$\pm$3\% of all the SDSS
QSO2s with detection in all the three shortest wavelength bands of
\wise fall in the infrared AGN wedge. This fraction increases to ${\rm
  66.1_{-4.7}^{+4.5}}\%$ at the highest \oiii {} luminosities, from
$\sim$2.2$\times10^{42}$ to 2.5$\times 10^{43}\,\lum$. This is
substantially lower than for the \buxs type 1 AGN in the same
luminosity range, but it is consistent within the uncertainties, with
that obtained for the \buxs type 2 AGN.

To evaluate whether our infrared selection technique identifies
Compton-thick AGN, we compiled a sample of SDSS QSO2s identified as
Compton-thick candidates in the literature. We found that the fraction
of SDSS QSO2s that are Compton-thick candidates and fall in the
infrared AGN wedge is consistent, within the uncertainties, with the
fraction of objects in the full sample that fall in that zone (${\rm
  66.7_{-18.5}^{+15.5}}\%$ versus ${\rm
  77.4_{-13.4}^{+10.4}}\%$). Furthermore, at the AGN luminosities
involved in the comparison, Compton-thick and Compton-thin SDSS QSO2s
have similar \wise colour distributions.

Considering only the objects in the infrared AGN wedge to minimize
host galaxy contamination, we found that the BUXS type 1 AGN have
bluer \colX{} colours than the type 2 AGN, and the SDSS QSO2s trace
the objects with the reddest colours. The SDSS QSO2s trace better than
X-ray selection the X-ray undetected \wise AGN candidates in the \buxs
survey area with the reddest colours in the wedge. These results
suggest a trend where higher (line of sight) absorption might be
associated with redder \colX{} colours.

To understand the nature and properties of the AGN candidates
identified with \wise we analysed the median UV-to-MIR rest-frame SEDs
of the AGN in BUXS and the SDSS QSO2s in the infrared AGN wedge. Both
an extincted AGN continuum and host galaxy stellar emission, are
required to fully account for the rest-frame UV/optical/near-infrared
SEDs of the \buxs type 2 AGN and SDSS QSO2s. Even for the SDSS QSO2s
the measured extinction is not sufficiently large to fully suppress
the nuclear emission at these wavelengths.

The SDSS QSO2s have more pronounced near-infrared stellar bumps than
both the type 1 and type 2 AGN in \nbuxs. The tight magnitude limits
applied by the SDSS spectroscopic target algorithms, especially at
\red$\leq$0.3, are likely biasing type 2 AGN samples towards objects
with more luminous (more massive) hosts. It does not appear that
either \red{} or luminosity effects can explain these differences.

The MIR continuum of both the \buxs AGN and SDSS QSO2s is dominated by
the AGN component. No clear features of star-forming activity were
detected in the MIR median SEDs of the \buxs type 1 and type 2
AGN. However MIR excess emission is a general property of the SDSS
QSO2s. We interpret this as features associated with star-forming
activity in the AGN host galaxies. The combination of extinction at
rest-frame optical/near-infrared wavelengths and contamination from
the more massive host galaxies at the longest MIR wavelengths
naturally explains the red \wise colours and MIR continua of the SDSS
QSO2s. Host galaxy dilution is an important effect at MIR wavelengths
that may strongly bias the AGN populations identified with \wise
colour-based techniques, especially at low luminosities, against AGN
in very massive/star-forming hosts.

We conclude that at high luminosities and \red$\lesssim$1 our MIR
selection technique is very efficient at identifying both Compton-thin
and Compton-thick AGN. Although optical/near-infrared spectroscopy is
essential to reveal the true nature of infrared-selected AGN, as
pointed out in M12, the best \wise candidates to account, at least in
part, for the absorbed luminous AGN missed in X-rays are the objects
with the reddest MIR colours in the wedge.

\section*{Acknowledgments}
This work is based on observations obtained with
{\textit{XMM-Newton}}, an ESA science mission with instruments and
contributions directly funded by ESA Member States and NASA. Based on
data from the Wide-field Infrared Survey Explorer, which is a joint
project of the University of California, Los Angeles, and the Jet
Propulsion Laboratory/California Institute of Technology, funded by
the National Aeronautics and Space Administration.  Funding for the
SDSS and SDSS-II has been provided by the Alfred P. Sloan Foundation,
the Participating Institutions, the National Science Foundation, the
U.S. Department of Energy, the National Aeronautics and Space
Administration, the Japanese Monbukagakusho, the Max Planck Society,
and the Higher Education Funding Council for England. The SDSS Web
Site is http://www.sdss.org/.  Based on observations collected at the
European Organisation for Astronomical Research in the Southern
Hemisphere, Chile, programme IDs 084.A-0828, 086.A-0612, 087.A-0447
and 088.A-0628. Based on observations made with the William Herschel
Telescope and its service programme -operated by the Isaac Newton
Group-, the Telescopio Nazionale Galileo -operated by the Centro
Galileo Galilei and the Gran Telescopio de Canarias installed in the
Spanish Observatorio del Roque de los Muchachos of the Instituto de
Astrofísica de Canarias, in the island of La Palma. SM and FJC
acknowledge financial support by the Spanish Ministry of Economy and
Competitiveness through grant AYA2010-21490-C02-01. SM acknowledge
financial support from the JAE-Doc program (Consejo Superior de
Investigaciones Científicas, cofunded by FSE). A.A.-H. acknowledges
support from the Universidad de Cantabria through the Augusto
G. Linares program. A.B. acknowledges a Royal Society Wolfson Research
Merit Award. P.S. acknowledges financial support from ASI (grant
No. I/009/10/0). A.R. acknowledges support from an IUCAA post-doctoral
fellowship. The authors thank A. Hern{\'a}n Caballero for useful
suggestions. The authors wish to thank the anonymous referee for
constructive comments.

\vspace{-0.2in}
\small
\bibliographystyle{mn2e}

\bsp

\label{lastpage}

\end{document}